\documentclass{aa}
\usepackage[varg]{txfonts}
\usepackage{graphicx}
\usepackage{amssymb}
\usepackage{natbib,twoopt}

\bibpunct{(}{)}{;}{a}{}{,} %% natbib format for A&A and ApJ
\usepackage{upgreek}
\usepackage{siunitx,booktabs,subcaption}
\usepackage{mathtools}
\usepackage{dcolumn}
\usepackage{amsmath}
\usepackage{color}
\usepackage{blindtext}
\usepackage{hyperref}
\hypersetup{
    colorlinks=true,
    linkcolor=blue,
    filecolor=blue,      
    urlcolor=blue,
    citecolor=blue,
}
\usepackage[labelfont={color=blue,bf}]{caption}
\begin{document}

\newcommand{\borrar}[1]{{\color{red} #1}}
\newcommand{\cme}[1]{{\color{cyan} #1}}
\newcommand{\rafa}[1]{{\color{magenta} #1}}
\newcommand{\ar}[1]{{\color{blue} #1}}

\title{Glitch time series and size distributions in eight prolific pulsars}

\author{J. R. Fuentes\inst{\ref{uc},\ref{mcgill}} \and C. M. Espinoza\inst{\ref{usach}}
\and A. Reisenegger\inst{\ref{uc}}
}

\institute{Instituto de Astrof\'isica, Pontificia Universidad Cat\'olica de Chile, Av. Vicu\~na Mackenna 4860, 7820436 Macul, Santiago, Chile \label{uc} \and 
Departamento de F\'isica, Universidad de Santiago de Chile, Avenida Ecuador 3493, 9170124 Estaci\'on Central, Santiago, Chile \label{usach} \and Department of Physics and McGill Space Institute, McGill University, 3600 rue University, H3A 2T8 Montreal QC, Canada\label{mcgill}
}

\date{Accepted XXX / Received YYY}

\abstract{ % CONTEXT
Glitches are rare spin-up events that punctuate the smooth slow-down of the rotation of pulsars.
For the Vela pulsar and PSR J0537$-$6910, the glitch sizes and the times between consecutive events have clear preferred scales (Gaussian distributions), contrary to the handful of other pulsars with enough glitches for such a study. Moreover, PSR J0537$-$6910 is the only pulsar showing a strong positive correlation between the size of each glitch and the waiting time until the following one.
}{ %AIMS
We attempt to understand this behaviour through a detailed study of the distributions and correlations of glitch properties for the eight pulsars with at least ten detected glitches.

}{ %METHODS
We model the distributions of glitch sizes and times between consecutive glitches for this sample.
Monte Carlo simulations are used to explore two hypotheses that could explain why the correlation is so much weaker in other pulsars than in PSR J0537$-$6910. 
}{ %RESULTS
We confirm the above results for the Vela pulsar and PSR J0537$-$6910, and verify that the latter is the only pulsar with a strong correlation between glitch size and waiting time to the following glitch.
For the remaining six pulsars, the waiting time distributions are best fitted by exponentials, and the size distributions either by power laws, exponentials, or log-normal functions. Some pulsars in the sample yield significant Pearson and Spearman coefficients ($r_p$ and $r_s$) for the aforementioned correlation. Moreover, for all except the Crab, both coefficients are positive. For each coefficient taken separately, the probability of this happening by chance is $1/16$. Our simulations show that the weaker correlations in pulsars other than PSR J0537$-$6910 cannot be due to missing glitches too small to be detected. We also tested the hypothesis that each pulsar may have two kinds of glitches, namely large, correlated ones and small, uncorrelated ones. The best results are obtained for the Vela pulsar, which exhibits a correlation with $r_p=0.68$ ($p$-value$=0.003$) if its 2 smallest glitches are removed. The other pulsars are harder to accommodate under this hypothesis, but their glitches are not consistent with a pure uncorrelated population either. We also found that all pulsars in our sample, except the Crab, are consistent with the previously found constant ratio between glitch activity and spin-down rate, $\dot\nu_g/|\dot\nu|=0.010\pm 0.001$, even though some of them have not shown any large glitches.
}
{ %CONCLUSIONS

To explain these results, we speculate that, except in the case of the Crab pulsar, all glitches draw their angular momentum from a common reservoir (presumably a neutron superfluid component containing $\approx 1\%$ of the star's moment of inertia), but two different trigger mechanisms could be active, a more deterministic one for larger glitches and a more random one for smaller ones.
}


\maketitle

\section{Introduction}
The rotation frequencies %rates 
$\nu$ of pulsars generally decrease slowly in time, but occasionally experience sudden increases $\Delta\nu$ that are usually accompanied by increases in the absolute value of their spin-down rates, $\dot{\nu}$ \citep{rm69,rd69,sl96}. 
These spin-up events, known as glitches, are infrequent, not periodic, and cover a wide range of sizes \citep[from $\Delta \nu/\nu \sim 10^{-11}$ to $\Delta \nu/\nu \sim 10^{-5}$;][]{elsk11,ymh+13}. 
The mechanism that generates these events is not completely understood, but they are believed to be caused by angular momentum transfer from an internal neutron superfluid %inside the neutron star 
to the rest of the neutron star \citep{ai75}.

Thanks to the few long-term monitoring campaigns that keep operating, some since the 1970s \citep[e.g.][]{hlk+04,ymh+13}, the number of detected glitches has slowly increased, thereby improving the significance of statistical studies in pulsar populations. 
\citet{ml90}, \citet{lsg00}, and \citet{elsk11} showed that the glitch activity $\dot{\nu}_{\rm{g}}$ (defined as the mean frequency increment per unit of time due to glitches) correlates linearly with $|\dot\nu|$. 
They also found that young pulsars (using the characteristic age, $\tau_c=-\nu/2\dot{\nu}$, as a proxy for age), which also have the highest $|\dot\nu|$, exhibit glitches more often than older pulsars, with rates varying from about one glitch per year to one per decade among the young pulsars. 
Using a larger and unbiased sample, \cite{fer+17} confirmed that the size distribution of all glitches in a large and representative sample of pulsars is multi-modal \citep[recently also seen by][]{ka14b,apj17}, with at least two well-defined classes of glitches: large glitches in a relatively narrow range $\Delta \nu \sim (10-30)\, \rm{\mu Hz}$, and small glitches with a much wider distribution, from $\sim 10\,\mathrm{\mu Hz}$ down to at least $10^{-4}\,\mathrm{\mu Hz}$. 
Further, \cite{fer+17} found that a constant ratio $\dot\nu_{\rm{g}}/|\dot\nu| = 0.010 \pm 0.001$ is consistent with the behaviour of nearly all rotation-powered pulsars and magnetars.
The only exception are the (few) very young pulsars, which have the highest spin-down rates, such as the Crab pulsar (PSR B0531$+$21) and PSR B0540$-$69. 

Because glitches are rare events, the number of known glitches in the vast majority of pulsars is not enough to perform robust statistical analyses on individual bases.
This has made people focus on the few objects that have the largest numbers of detected glitches (about 10 pulsars).
The statistical distributions of glitch sizes and times between consecutive glitches (waiting times), for the nine pulsars with more than five known glitches at the time, were studied by \citet{mpw08}.
They found that seven out of the nine pulsars exhibited power-law-like size distributions and exponential waiting time distributions. %, which made them conclude that glitches in those pulsars are produced by avalanche processes\rafa{preferiria no mencionar las avalanchas}.
The distributions of the other two (PSRs J0537$-$6910 and B0833$-$45, the Vela pulsar) were
%However, they also found two exceptions for which both distributions were 
better described by Gaussian functions, setting preferred sizes and time scales.
These results have been further confirmed by \citet{fmh17} and \citet{hmd18}, who also found that there are at least two main behaviours among the glitching pulsars.

Correlations between glitch sizes and the times to the nearest glitches, either backward or forward, are naturally expected.
We know that glitch activity is driven by the spin-down rate \citep{fer+17}, which suggests that glitches are the release of some stress that builds up  at a rate determined by $|\dot{\nu}|$.
If the stress is completely released at each glitch, then one should expect a correlation between size and the time since the last glitch.
Conversely, if glitches occur when a certain critical state is reached, one should expect a correlation between size and the time to the next glitch, as longer times would be needed to come back to the critical state after the largest glitches.
Moreover, if both assumptions were indeed correct, glitches would all be of equal sizes and occur periodically. 
However, with the exception of PSR J0537$-$6910 (see below), no other pulsars have shown significant correlations between glitch sizes and the times to the nearest events \citep[e.g.][]{wmp+00,ywml10,mhf18}.
This may be partly due to small-number statistics and might improve in the future, provided a substantial number of pulsars continue to be monitored for glitches.

The case of PSR J0537$-$6910, however, is very clear.
With more than 40 glitches detected in $\sim 13$\,yr, the statistical conclusions about its behaviour are much more significant than for any other pulsar.
As first reported by \citet{mmw+06}, its glitch sizes exhibit a strong correlation with the waiting time to the following glitch \citep[see also][who confirmed the correlation using twice as much data]{aeka18,fagk18}.

\citet{aeka18} interpret %such 
this behaviour as an indication that glitches in this pulsar occur only once some threshold is reached. %, and showed that the evolution of the system towards this threshold is driven by the spin-down.}
Moreover, this behaviour would imply that not necessarily all the stress is released in the glitches, thereby giving rise to the variety of (unpredictable) glitch sizes observed and the lack of backward time correlation.

In this work we study the sequence of glitches in the pulsars with at least ten detected events, by characterizing their distributions of glitch sizes and waiting times between successive glitches. Also, we test two hypotheses to explain
why most pulsars do not show a correlation between glitch size and time to the following glitch: the effects of undetected small glitches and the possibility that two different classes of glitches are present in each pulsar.

\section{Pulsars with at least ten detected glitches} 
\label{s1}

\begin{figure}
\centering
\includegraphics[width=9cm]{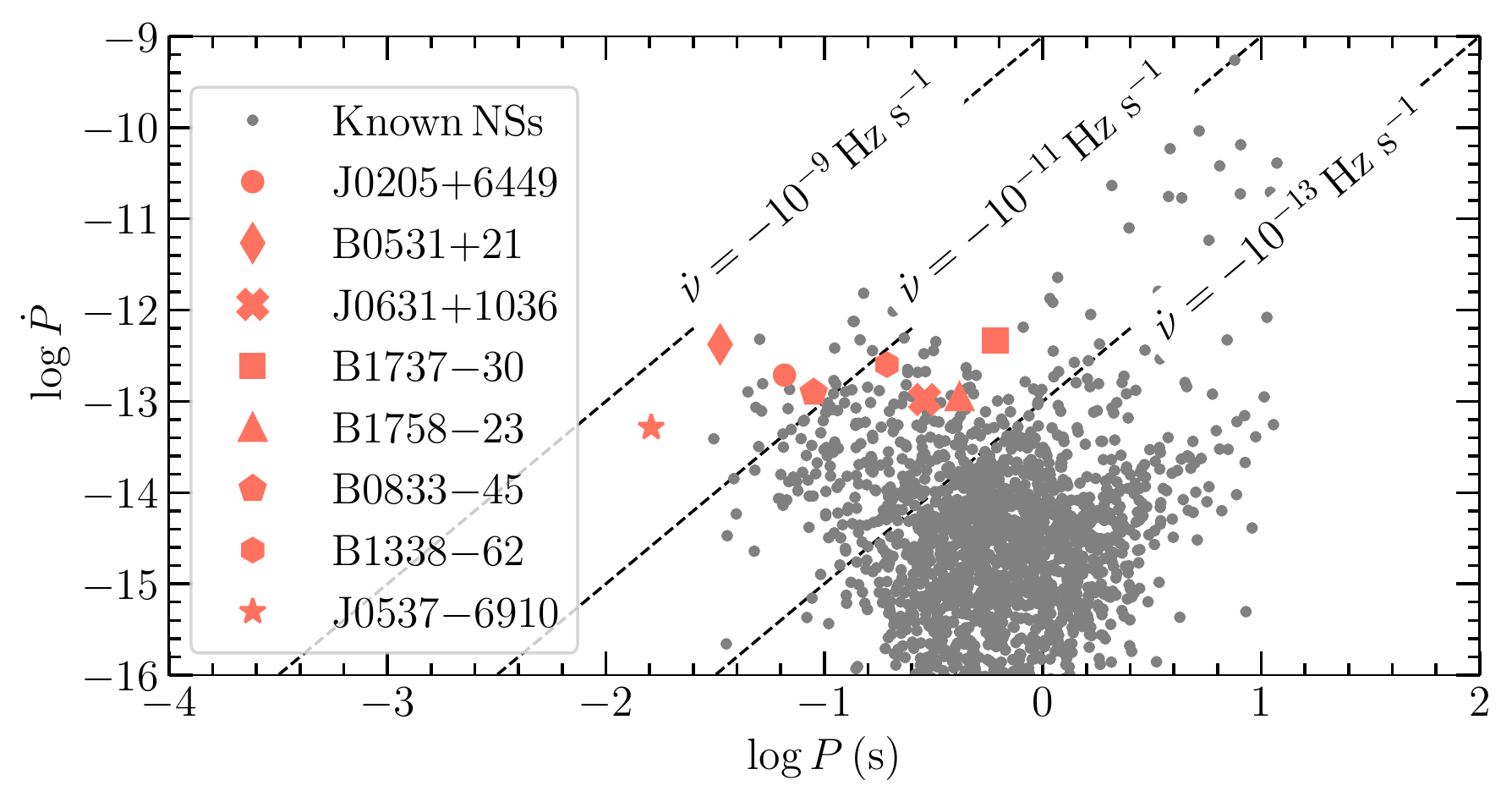}
\caption{Upper part of the $P-\dot{P}$ diagram for all known pulsars. 
	The pulsars in our sample have at least ten detected glitches and are labeled with different symbols.  
	Lines of constant spin-down rate $\dot{\nu}$ are shown and labeled. 
	$P$ and $\dot{P}$ values were taken from the ATNF pulsar catalog \protect\footnotemark.
}\label{fig1}
\end{figure}
\footnotetext{\url{http://www.atnf.csiro.au/research/pulsar/psrcat}}
To date, there are eight pulsars with at least 10 detected glitches (Fig. \ref{fig1}).
PSRs J0205$+$6449, B0531$+$21 (the Crab pulsar), B1737$-$30, B1758$-$23, and J0631$+$1036 have been observed regularly by the Jodrell Bank Observatory \citep[JBO,][]{hlk+04}.
PSR B1338$-$62 has been observed by the Parkes telescope, and  the Vela pulsar has been observed by several telescopes, including Parkes, the Jet Propulsion Laboratory, and others in Australia and  Southafrica \citep[e.g.][]{downs81,mkhr87,ymh+13,buc13}. 
PSR J0537$-$6910 is the only object in our sample not detected in the radio band and was observed for 13 years by the \textit{Rossi X-ray Timing Explorer} \citep[RXTE,][]{aeka18,fagk18}. 
Glitch epochs and sizes were taken from the JBO online glitch catalog \footnote{\url{http://www.jb.man.ac.uk/pulsar/glitches/gTable.html}}, where more information and the appropriate references for each measurement can be found.

\begin{figure*}
\centering
\includegraphics[width=18cm]{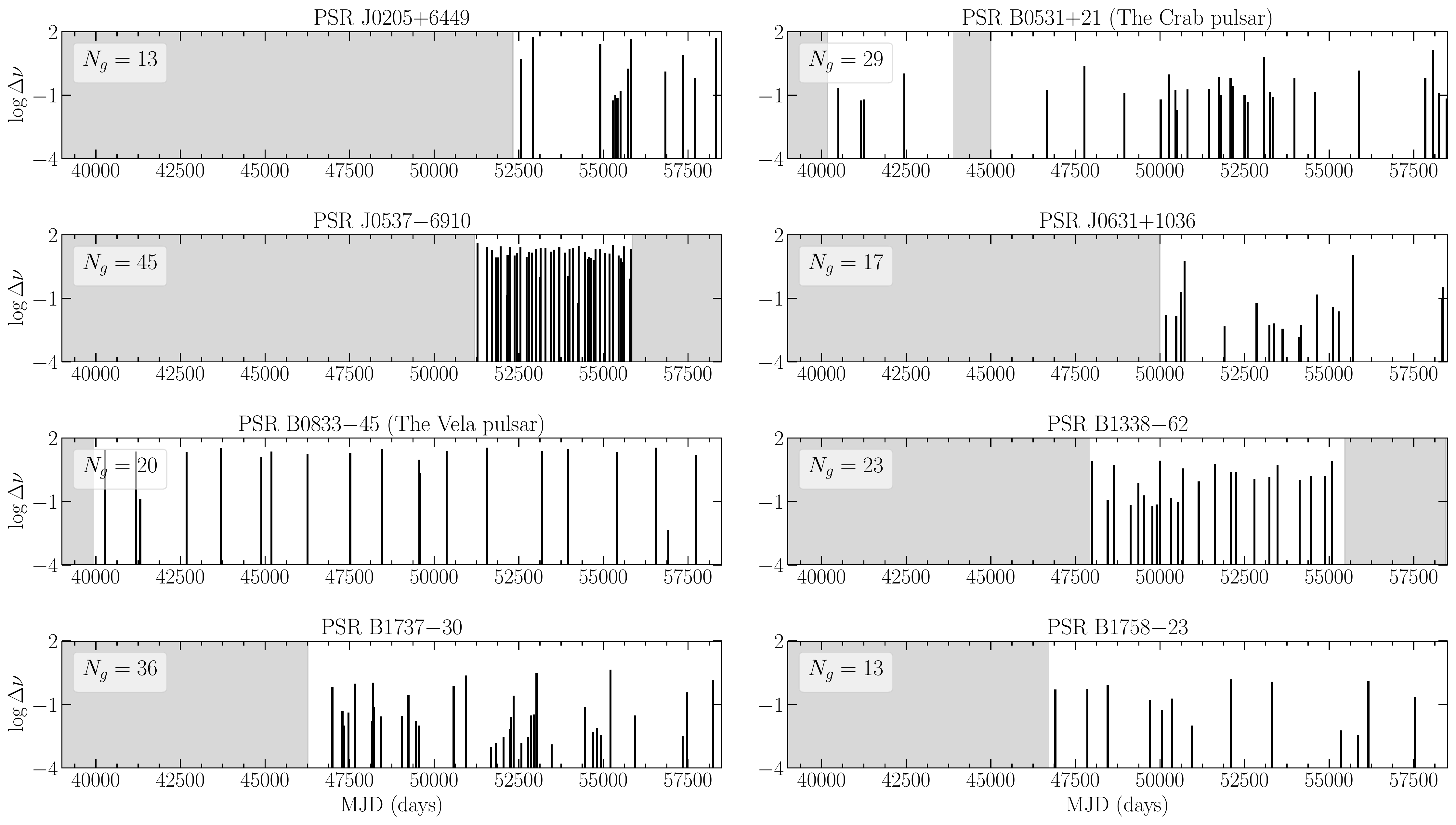}
\caption{Logarithm (base 10) of glitch sizes $\Delta\nu$ (with $\Delta\nu$ measured in $\mu$Hz) as a function of the glitch epoch for the pulsars in the sample. 
	The gray areas mark periods of time in which there were no observations for more than 3 months. 
	$N_g$ is the number of glitches detected in the respective pulsar, until 20 April 2019 (MJD 58593). 
	To build a continuous sample, in the analyses of the Crab pulsar, we only use the 25 glitches after MJD 45000, when daily observations started \citep{eas+14}. All panels share the same scale, in both axes. %The total time span corresponds to $53.25$\,yr.
	} \label{fig2}
\end{figure*}

\begin{figure*}
\centering
\includegraphics[width=18cm]{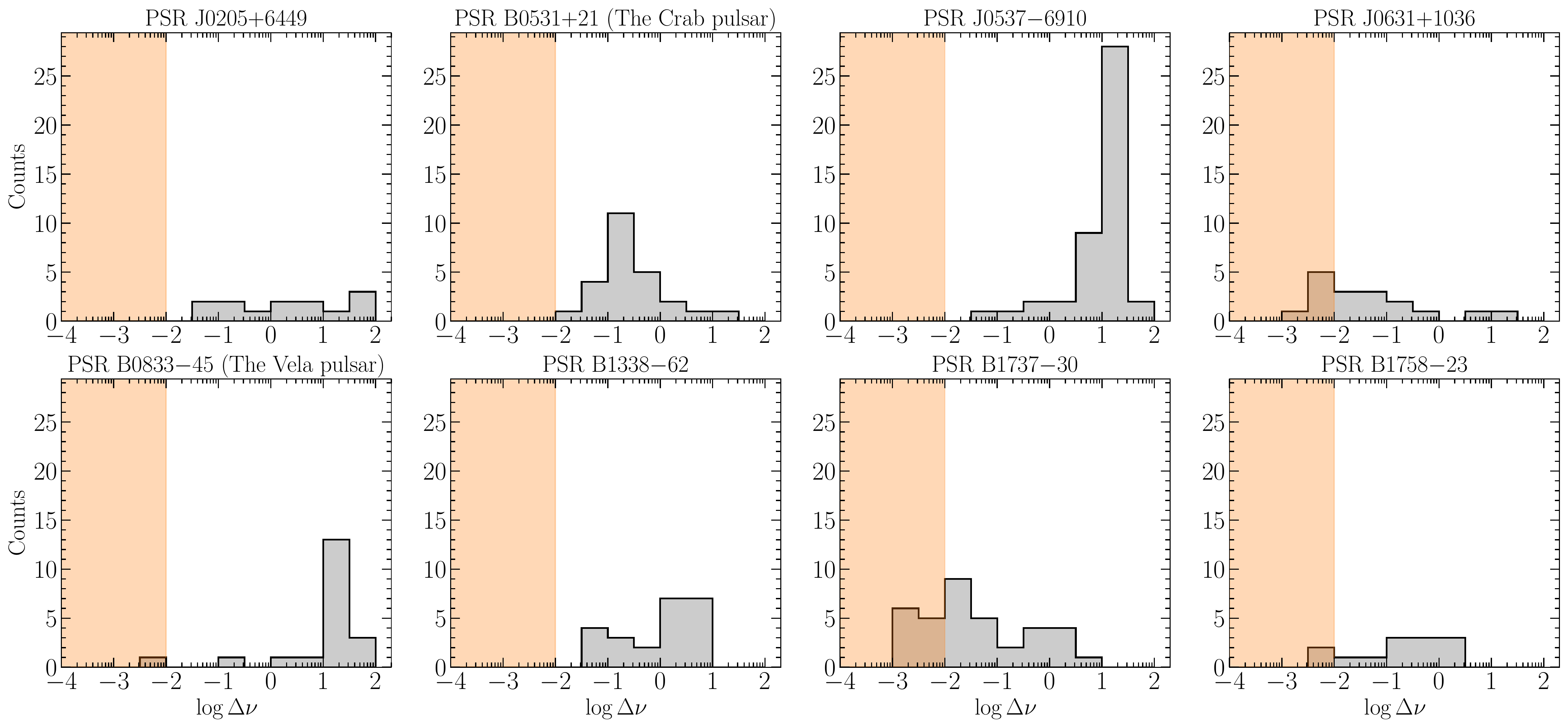}
\caption{Distribution of $\log \Delta \nu$ (with $\Delta \nu$ measured in $\rm{\mu Hz}$) for the pulsars in our sample. The orange areas indicate that glitches with $\Delta\nu<0.01\,\rm{\mu Hz}$ could be missing due to detectability issues.} \label{fig3}
\end{figure*}

Figures \ref{fig2} and \ref{fig3} show that the Vela pulsar and PSR J0537$-$6910 produce glitches of similar sizes, particularly large glitches ($\Delta \nu > 10$ $\mu$Hz), and in fairly regular time intervals. The absence of smaller glitches in these pulsars is not a selection effect, as it is quite unlikely that a considerable amount of glitches with sizes up to $\Delta \nu \sim 10\,\rm{\mu Hz}$, far above the detection limits reported in the literature \citep[see ][and text below]{wxe+15}, could have gone undetected. 
On the other hand, the rest of the pulsars exhibit irregular waiting times and cover a wider range of sizes ($\Delta \nu \sim 10^{-3}-10$ $\mu$Hz). 
%PSR J0205$+$6449, with more limited statistics (only 13 glitches detected), also has mainly large glitches and it is the pulsar with the largest glitch detected so far among all known pulsars ($\Delta\nu = 58\, \mu$Hz).

The cadence of the timing observations varies considerably from pulsar to pulsar (and even with time for individual pulsars), and the sensitivity of the  observations, from which the glitch measurements were performed, are also different between different pulsars.
This means that the chances of detecting very small glitches are different for each pulsar and that the completeness of the samples towards small events might also be different \citep{eas+14}.
Nonetheless, in this study we use a single value to represent the glitch size below which samples are likely to be incomplete due to detectability issues.
For an observing cadence of 30 days and a rotational noise of 0.01 rotational phases, glitch detection is severely compromised below sizes $\Delta\nu \sim 10^{-2}\, \rm{\mu Hz}$, especially if their frequency derivative steps are larger than $|\Delta\dot{\nu}|\sim 10^{-15} \, \rm{Hz\, s^{-1}}$ \citep[see][]{wxe+15}. We use the above numbers to characterize the glitch detection capabilities in this sample of pulsars, but we note that such cadence and rotational noise are rather pessimistic values in some cases.

\section{Distributions of glitch sizes and times between glitches} 
\label{distris}
 
In the following, we model the distributions of glitch sizes ($\Delta\nu$, measured in $\mu$Hz) and the distributions of times between successive glitches ($\Delta \tau$, measured in yr) for each pulsar in our sample. 
Four probability density distributions are considered: Gaussian,
\begin{equation}
M(x|\mu,\sigma) = C_{\rm{Gauss}}\,\exp\left[\frac{-(x-\mu)^2}{2\sigma^2}\right]\text{,}
\end{equation}
power-law,
\begin{equation}
M(x|\alpha) = \dfrac{\alpha - 1}{x_{\rm{min}}}\left(\dfrac{x}{x_{\rm{min}}}\right)^{-\alpha}\text{,}
\end{equation}
log-normal,
\begin{equation}
M(x|\mu_{\rm{L-N}},\sigma_{\rm{L-N}}) = \dfrac{C_{\rm{L-N}}}{x}\,\exp\left[\frac{-(\ln x-\mu_{\rm{L-N}})^2}{2\sigma_{\rm{L-N}}^2}\right]\text{,}
\end{equation}
and exponential,
\begin{equation}
M(x|\lambda) = \lambda\, \exp\left[-\lambda(x-x_{\rm{min}})\right]\text{.}
\end{equation}

The set $\{\mu,\sigma, \alpha, \mu_{\rm{L-N}},\sigma_{\rm{L-N}}, \lambda\}$ are the fitting parameters.  
All the distributions are normalized in the range $x_{\rm{min}}$ to $\infty$. Formally, $x_{\rm{min}}$ is given by detection limits. 
However, it is not simple to define precise values for $\Delta \nu_{\rm{min}}$ and $\Delta \uptau_{\rm{min}}$ for each pulsar.
Thus we use $\Delta \nu_{\rm{min}} = 10^{-2}\, \mu$Hz for the glitch sizes (see previous section), and  the smallest interval of time between glitches in each pulsar as $\Delta \uptau_{\rm{min}}$.

For the Gaussian and log-normal distributions the normalization constants $C_{\rm{Gauss}}$ and $C_{\rm{L-N}}$ were found numerically.
We use the maximum likelihood technique to obtain the parameters of the models that describe best the data, and use the Akaike Information Criterion \citep[AIC,][]{aka74} to compare the different models \citep[see also the Appendix in][]{fer+17}.

\begin{figure*}
\includegraphics[width=18cm]{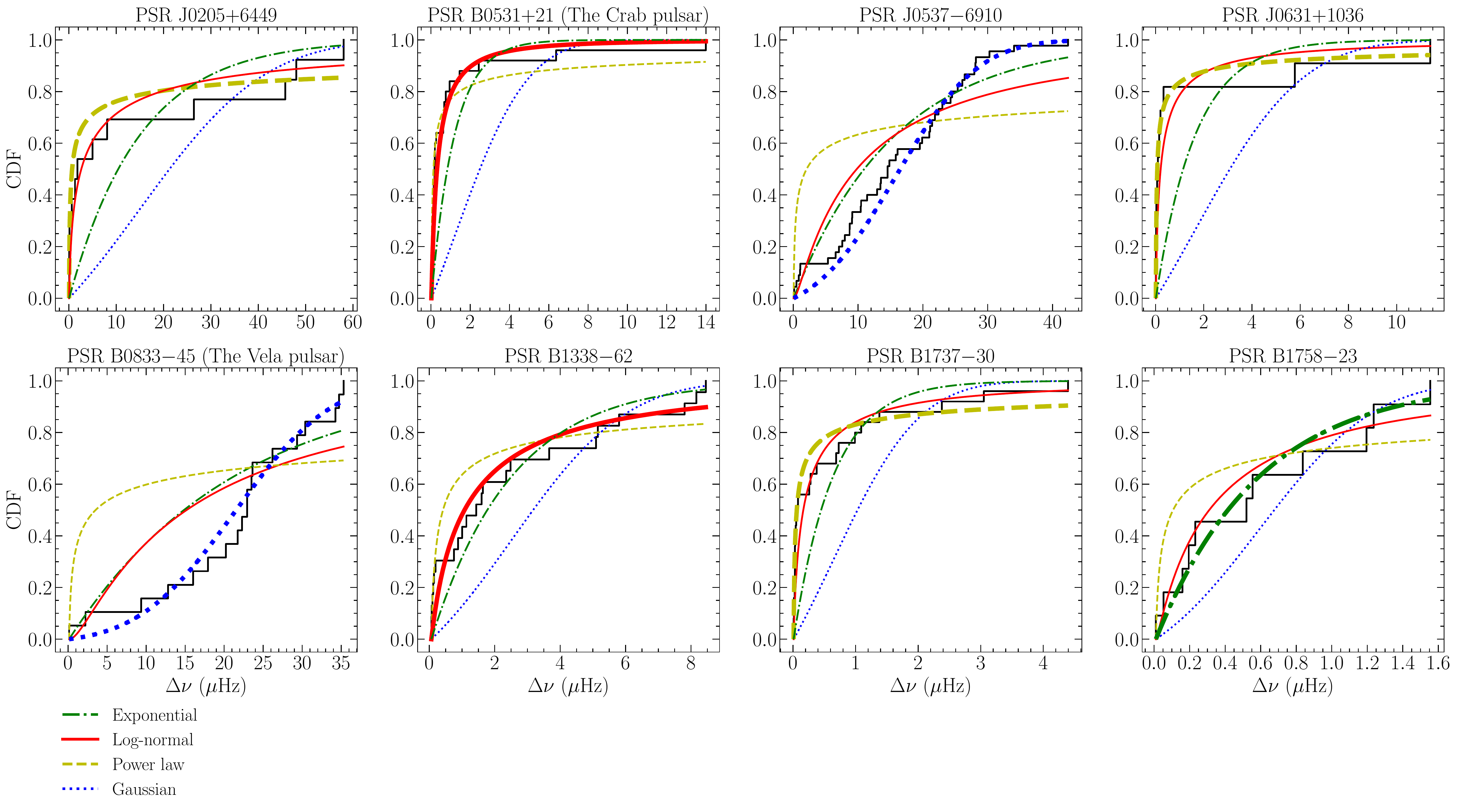}
\caption{Cumulative distribution of glitch sizes and model fits. The best-fitting models are indicated by thicker curves.}\label{fig4}
\end{figure*}

\begin{figure*}
\includegraphics[width=18cm]{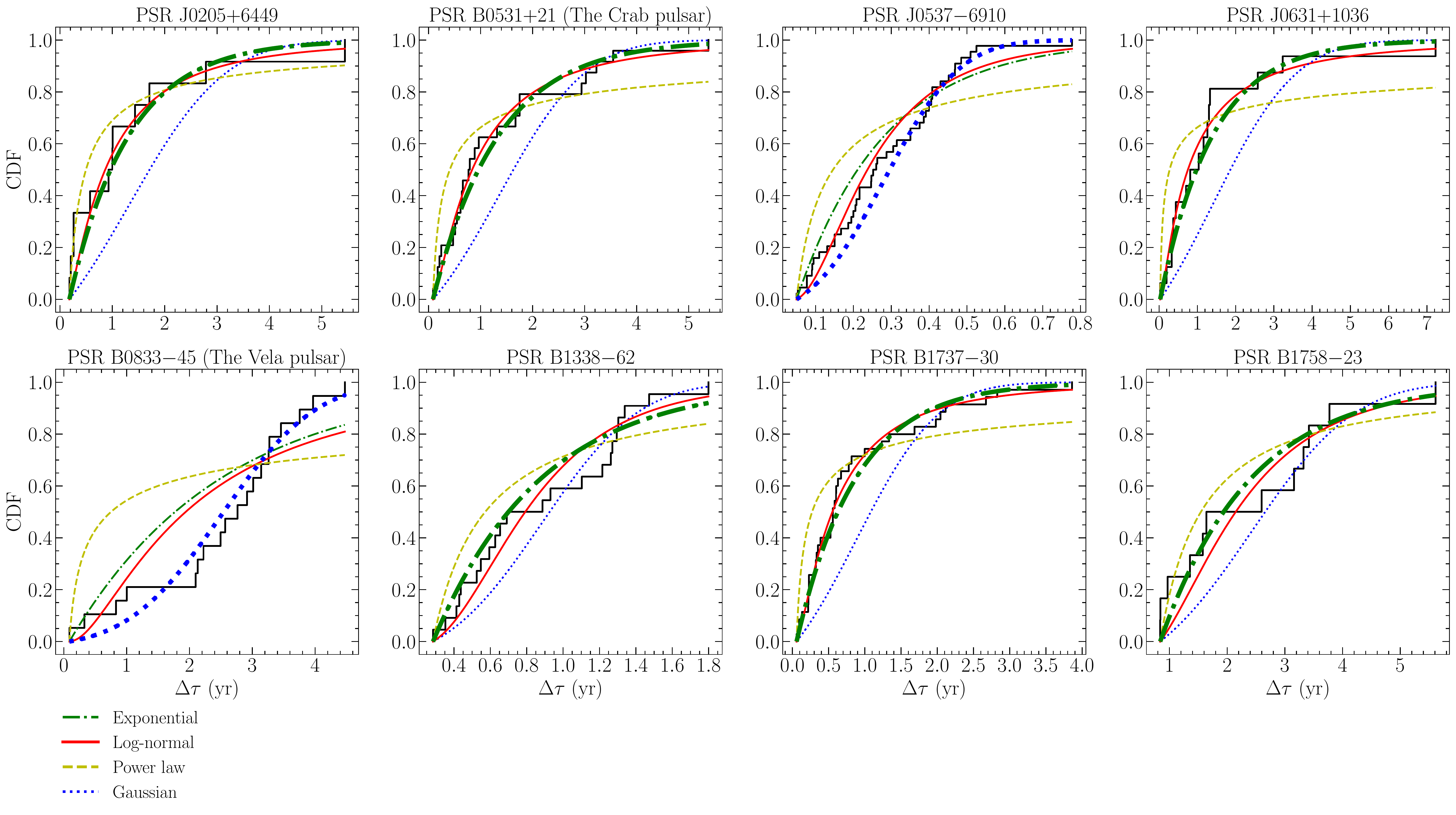}
\caption{Cumulative distribution of waiting times between successive glitches and model fits. The best-fitting models are indicated by thicker curves.}\label{fig5}
\end{figure*}

\begin{table*}%[!ht]
\centering
\caption{Distributions of glitch sizes: results of the fits and the AIC weights for each model; using glitches with $\Delta\nu \geq 0.01\, \rm{\mu Hz}$.}
\label{Table_sizes}
\begin{tabular}{@{}lcccccccccc@{}}
\toprule \toprule
PSR Name & $w^{\rm{Gauss}}$ & $w^{\textrm{Power law}}$ & $w^{\textrm{L-N}}$ & $w^{\rm{Exp}}$ & $\hat{\mu}$ & $\hat{\sigma}$ & $\hat{\alpha}$ & $\hat{\mu}_{\rm{L-N}}$ & $\hat{\sigma}_{\rm{L-N}}$ & $\hat{\lambda}$ \\
 & & & & & $\rm{\mu Hz}$ & $\rm{\mu Hz}$ &  &  &  & $(\rm{\mu Hz})^{-1}$\\
\midrule
J0205$+$6449 & $10^{-8}$ & $\mathbf{0.66}$ & 0.33 & $10^{-5}$ & 15(5) & 20(4) & 1.27(6) & 0.7(7) & 2.5(3) & 0.07(6)\\

B0531$+$21 & $10^{-17}$ & 0.02 & $\mathbf{0.97}$ & $10^{-7}$ & 1.2(5) & 3(1) & 1.4(1) & -1.3(3) & 1.5(2) & 0.8(7)\\

J0537$-$6910 & $\mathbf{0.96}$ & $10^{-24}$ & $10^{-8}$ & 0.03 & 15(1) & 9.9(9) & 1.19(5) & 2.2(2) & 1.3(2) & 0.063(6)\\

J0631$+$1036 & $10^{-12}$ & $\mathbf{0.94}$ & 0.05 & $10^{-8}$ & 1(1) & 3(1) & 1.4(1) & -1.9(6) & 2.1(4) & 0.61(4)\\

B0833$-$45 & $\mathbf{0.997}$ & $10^{-13}$ & $10^{-6}$ & 0.002 & 21(2) & 9(1) & 1.2(4) & 2.7(2) & 1.2(4) & 0.05(1)\\

B1338$-$62 & $10^{-5}$ & 0.07 & $\mathbf{0.53}$ & 0.4 & 2.5(5) & 2.7(3) & 1.36(5) & -0.1(3) & 1.6(1) & 0.4(1)\\

B1737$-$30 & $10^{-14}$ & $\mathbf{0.82}$ & 0.17 & $10^{-7}$ & 0.6(2) & 1.0(2) & 1.38(6) & -2.0(3) & 1.9(1) & 1.5(8)\\

B1758$-$23 & 0.06 & 0.004 & 0.07 & $\mathbf{0.866}$ & 0.6(1) & 0.51(8) & 1.3(2) & -1.2(4) & 1.5(3) & 1.7(6)\\
\bottomrule
\end{tabular}
\tablefoot{$w^m$ denotes the Akaike weight of the model $m$. 
$\hat \mu$ and $\hat \sigma$ are the mean and the standard deviation of the Gaussian model, and $\hat\alpha$ is the power-law index. 
$\hat \mu_{\rm{L-N}}$ and $\hat \sigma_{\rm{L-N}}$ are the mean and the standard deviation of the log-normal model, respectively. $\hat \lambda$ is the rate parameter of the exponential distribution. 
The values in parentheses correspond to the uncertainty in the last quoted digit and were calculated using the usual bootstrap method. We marked in bold the values of $w^m$ for the best models.}
\end{table*}

\begin{table*}
\centering
\caption{Distributions of waiting times between successive glitches: results of the fits and the AIC weights for each model.}\label{Table_times}
\begin{tabular}{@{}lcccccccccc@{}}
\toprule \toprule
PSR Name & $w^{\rm{Gauss}}$ & $w^{\rm{Power law}}$ & $w^{\rm{L-N}}$ & $w^{\rm{Exp}}$ & $\hat{\mu}$ & $\hat{\sigma}$ & $\hat{\alpha}$ & $\hat{\mu}_{\rm{L-N}}$ & $\hat{\sigma}_{\rm{L-N}}$ & $\hat{\lambda}$\\
& & & & & yr & yr &  &  &  & yr$^{-1}$\\
\midrule

J0205$+$6449 & $0.001$ & $0.40$ & 0.16 & $\mathbf{0.43}$ & 1.3(4)& 1.4(4) & 1.7(1) & -$0.2(3)$ & 1.0(1) & 0.9(5)\\

B0531$+$21 & $10^{-4}$ & $10^{-5}$ & 0.15 & $\mathbf{0.84}$ & 1.3(2) & 1.3(2) & 1.4(1) & -$0.2(2)$ & 1.0(1) & 0.8(2)\\

J0537$-$6910 & $\mathbf{0.72}$ & $10^{-10}$ & 0.07 & 0.2 & 0.28(2) & 0.15(1) & 1.64(8) & -1.44(9) & 0.65(6) & 4.3(4)\\

J0631$+$1036  & $10^{-4}$ & $10^{-5}$ & 0.20 & $\mathbf{0.79}$ & 1.4(4) & 1.7(6) & 1.3(2) & -0.3(3) & 1.2(2) & 0.7(3)\\

B0833$-$45  & $\mathbf{0.993}$ & $10^{-10}$ & $10^{-4}$ & 0.006 & 2.5(2) & 1.2(1) & 1.3(3) & 0.7(2) & 0.9(2) & 0.41(9)\\

B1338$-$62  & 0.25 & $10^{-3}$ & 0.20 & $\mathbf{0.54}$ & 0.88(9) & 0.42(4) & 1.9(2) & -0.3(1) & 0.51(5) & 1.7(3)\\

B1737$-$30  & $10^{-5}$ & $10^{-6}$ & 0.17 & $\mathbf{0.82}$ & 0.9(1) & 0.9(1) & 1.44(7) & -0.6(1) & 1.0(1) & 1.2(2)\\

B1758$-$23  & 0.04 & 0.16 & 0.08 & $\mathbf{0.72}$ & 2.4(4) & 1.4(2) & 2.1(2) & 0.7(1) & 0.61(8) & 0.6(2)\\
\bottomrule
\end{tabular}
\tablefoot{$w^m$ denotes the Akaike weights of the model $m$. $\hat \mu$ and $\hat \sigma$ are the mean and the standard deviation of the Gaussian model, and $\hat\alpha$ is the power-law index. $\hat \mu_{\rm{L-N}}$ and $\hat \sigma_{\rm{L-N}}$ are the mean and the standard deviation of the log-normal model, respectively.  $\hat \lambda$ is the rate parameter of the exponential distribution. The values in parentheses correspond to the uncertainties in the last digit, and were calculated by using the bootstrap method. We marked in bold the values of $w^m$ for the best models.}
\end{table*}

Figures \ref{fig4}-\ref{fig5} and Tables \ref{Table_sizes}-\ref{Table_times} summarize the results of fitting these distributions to each pulsar. There is no single distribution type that can simultaneously describe all the pulsars satisfactorily, for either sizes or waiting times.
The size distributions present a large variety (as also found in the model of \citealt{cm19}): the log-normal distribution gives the best fit for the Crab pulsar and PSR B1338$-$62, power-law for PSRs J0631$+$1036, B1737$-$30, and J0205$+$6449, and exponential for PSRs B1758$-$23.

%Creo que se podia mal interpretar este parrafo, ya que leyendo en detalle los estudios previos, ellos solo testearon power-law y gaussiana, por lo que no hay muchas opciones para comparar. Asi que preferiria sacarlo. We should mention that for the Crab pulsar our results differ to what other studies have concluded \citep[e.g.][]{mpw08,hmd18,sls+18}, who found that a power-law gives the best fit. However, a power-law distribution is not consistent with the lack of small glitches below $\Delta \nu = 10^{-2}\, \mu$Hz reported by \citet{eas+14}. Furthermore, \citet{eas+14} found that a KS test gives similar results for both, power-law and log-normal distributions.

We also note that PSR J0205+6449 and PSR B1758$-$23 are the pulsars with the fewest recorded glitches in the sample (both have 13 glitches detected), hence we ought to wait and confirm this result once more events are detected.

In the case of PSRs J0537$-$6910 and B0833$-$45 (Vela), the best fit for both size and waiting time distributions are Gaussian functions.
Their size distributions are centered at large sizes $\Delta \nu \approx 15$  and $20\, \rm{\mu Hz}$, respectively, consistent with the peak of large glitches in the combined distribution for all pulsars \citep{fer+17}. 
%A Gaussian distribution for their waiting times gives support to the evidence for quasi-periodicity reported by \cite{mpw08} and \cite{hmd18}.

The distributions of times between successive glitches offer more homogeneous results. Besides the case of PSR J0537$-$6910 and the Vela pulsar (best modelled by Gaussian functions), the waiting time distributions for all the other pulsars are best represented by exponential functions.
These results are in agreement with \citet{mpw08,wwty12}, and \citet{hmd18} for almost all the pulsars studied. 
The only exception is PSR B1338$-$62, for which \cite{hmd18} reported a local maximum in the distribution and classified this pulsar as a quasi-periodic glitcher.

If $\Delta\nu_{\rm{min}}$ is set to the size of the smallest detected glitch in each pulsar (rather than to $10^{-2}\, \mu$Hz), the results of the fits are very similar, and give parameters within the uncertainties presented in Table \ref{Table_sizes}.

\section{Time series correlations: Glitch size and time to the next glitch} 
\label{s2}

\begin{figure*}
\centering
\includegraphics[width=18cm]{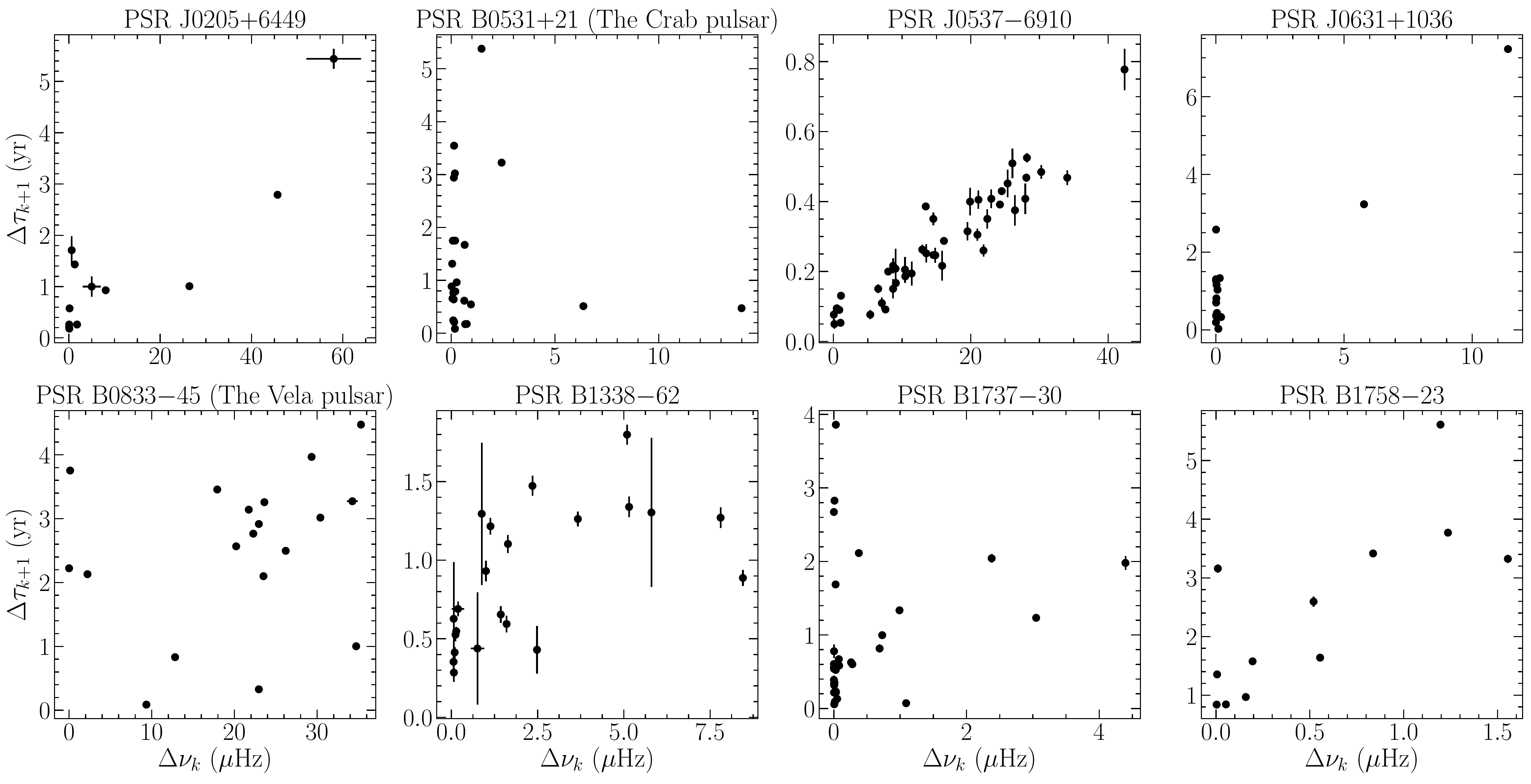}
\caption{Time to next glitch, $\Delta \uptau_{k+1}$, as a function of glitch size, $\Delta \nu_k$, for all the pulsars in the sample.} 
\label{fig6}
\end{figure*}

Different studies have shown that for PSR J0537$-$6910 the glitch magnitudes $\Delta \nu_k$ are strongly correlated with the waiting times to the following glitch $\Delta \uptau_{k+1}$ \citep[][and see Fig. \ref{fig6}]{mmw+06,aeka18,fagk18}. 

We test whether this correlation is also present in the other pulsars of the sample, and show the results in Table \ref{dnu_dt_next} and Fig. \ref{fig6} \citep[this is fairly consistent with][though we note that the samples of glitches are not exactly the same]{mhf18}. None of them exhibits a correlation as clear as PSR J0537$-$6910. 
However, for PSRs J0205$+$6449, J0631$+$1036, B1338$-$62, and B1758$-$23, the Pearson correlation coefficients are larger than $0.5$ and the $p$-values are $\sim 10^{-3}$, or less. 
Therefore, at $95\%$ confidence level ($p$-values $ < 0.05$), we can reject the null hypothesis that $\Delta \nu_k$ and $\Delta \uptau_{k+1}$ are uncorrelated in these pulsars.
Since the Pearson coefficient can be dominated by outliers, we also compute the Spearman rank correlation coefficient, obtaining similar or even stronger correlations, except for PSR J0631$+$1036.

\begin{table}
\centering
\caption{Correlation coefficients between $\Delta \nu_k$ and $\Delta \uptau_{k+1}$.} 
\begin{tabular}{@{}lccccc@{}}
\toprule \toprule
PSR Name & $N_{\mathrm{g}}$ & $r_p$ & $p_p$ & $r_s$ & $p_s$ \\
\midrule
J0205$+$6449 & 13 & 0.88 & 0.0002 & $0.76$ & 0.004 \\
B0531$+$21 & 25 & -0.10 & 0.62 & -0.12 & 0.57 \\
J0537$-$6910 & 45 & 0.95 & $10^{-22}$  & 0.95 & $10^{-23}$ \\
J0631$+$1036 & 17 & 0.93 & $10^{-7}$ & 0.20 & 0.45 \\
B0833$-$45 & 20 & 0.24 & 0.31 & 0.31 & 0.21 \\
B1338$-$62 & 23 & 0.59 & 0.003 & 0.70 & 0.0002\\
B1737$-$30 & 36 & 0.29 & 0.09 & 0.29 & 0.08 \\
B1758$-$23 & 13 & 0.76 & 0.003 & 0.80 & 0.001 \\
\bottomrule
\end{tabular}
\tablefoot{The first and second columns contain the names of the pulsars and the respective number of glitches detected, respectively. 
The third and fourth columns correspond to the Pearson linear correlation coefficient $r_p$ and the respective $p$-value $p_p$. 
The last two columns correspond to the Spearman correlation coefficient $r_s$ and the respective $p$-value $p_s$.
%\cme{0205: A MI ME DA $r_p=0.88$, $p_p=0.0002$, AL REDONDEAR.}
	}\label{dnu_dt_next}
\end{table}

It is also interesting to note that not only for PSR J0537$-$6910, but for all pulsars in the sample except the Crab, both the Pearson and Spearman correlation coefficients are positive. 
The probability of finding at least six out of seven pulsars having the same sign as our reference case, just by chance, is rather low. 
The probability of getting exactly $k$ successes among $n$ trials, with $1/2$ success probability in each trial, is $P(k\,|\,n) = {n\choose k}(1/2)^n$. 
Thus, the probability of getting at least 6 successes in 7 trials is

\begin{equation}
P(\geq 6\,|\,7)=P(6\,|\,7)+P(7\,|\,7)=\frac{1}{16}=0.0625\,.
\end{equation}

This low probability suggests that the waiting time to the following glitch is at least partially regulated by the size of the previous glitch.

In order to explain why the correlation for all other pulsars is much less clear than
for PSR J0537$-$6910, we explore two hypotheses, both of which are motivated by noting that most glitches in PSR J0537$-$6910 are large:

\begin{itemize}
\item[(I)] The correlation is intrinsically present in the full population of glitches of each pulsar, but glitches below a certain size threshold are not detected, thereby increasing by random amounts the times between the detected ones and worsening the correlation.\\

\item[(II)] There are two classes of glitches: glitches above a certain threshold size that follow the correlation, and glitches below the same threshold that are uncorrelated. 

\end{itemize}

\subsection{Hypothesis I: Incompleteness of the sample}

In order to test the first hypothesis, we simulate a hypothetical pulsar with 100 glitches that follow a perfect correlation between $\Delta\nu_{k}$ and $\Delta \uptau_{k+1}$.
The events smaller than a certain value are then removed to understand the effect of their absence in the correlation. 
The procedure is the following:

\begin{enumerate}

\item
Glitch sizes are generated from a power-law distribution given by $dN/d \Delta\nu \propto \Delta\nu^{-\alpha}$, with power-law index $\alpha>1$. 
We choose a power-law distribution because it mainly produces small events, and we want to see the effect of removing a substantial fraction of them.
Several different choices for $\alpha$ were considered. 
Here we only show the results for $\alpha = 1.2$ and 1.4, as they generate distributions that resemble some of the ones observed.

The distributions do not have an upper cutoff, and the lower limit was varied so that, after reducing the sample of glitches (as we explain in step 3 below), the resulting sample covers the typical observed range of glitch sizes ($10^{-2} - 10^2\, \rm{\mu Hz}$).

\item
The time to the next glitch $\Delta \uptau_{k+1}$ is computed in terms of the glitch size $\Delta \nu_k$ as:
\begin{equation}
\Delta\uptau_{k+1}= C\Delta\nu_{k}\, . 
\label{eq_corr}
\end{equation}
The value of the proportionality constant $C$ is irrelevant in this case, since we are simulating a generic pulsar.
\\

\item 
Steps (1) and (2) are repeated until a sequence of 100 glitches is reached. 
Then the 80 smallest are removed, thereby leaving a reduced sample of 20 to be analyzed, which is comparable to the number of glitches observed in each of our 8 pulsars.
The lower limit for the distribution is computed analytically so that, after reducing the sample of glitches, the final sample covers the typical observed range of glitch sizes ($10^{-2} - 10^2\, \rm{\mu Hz}$).\\

\item 
Finally, we calculate the time interval between each pair of successive glitches in the reduced sample,  and determine both the Spearman and Pearson correlation coefficients between $\Delta \nu_k$ and $\Delta \uptau_{k+1}$.

\end{enumerate}

\begin{figure}
\centering
\includegraphics[width=8cm]{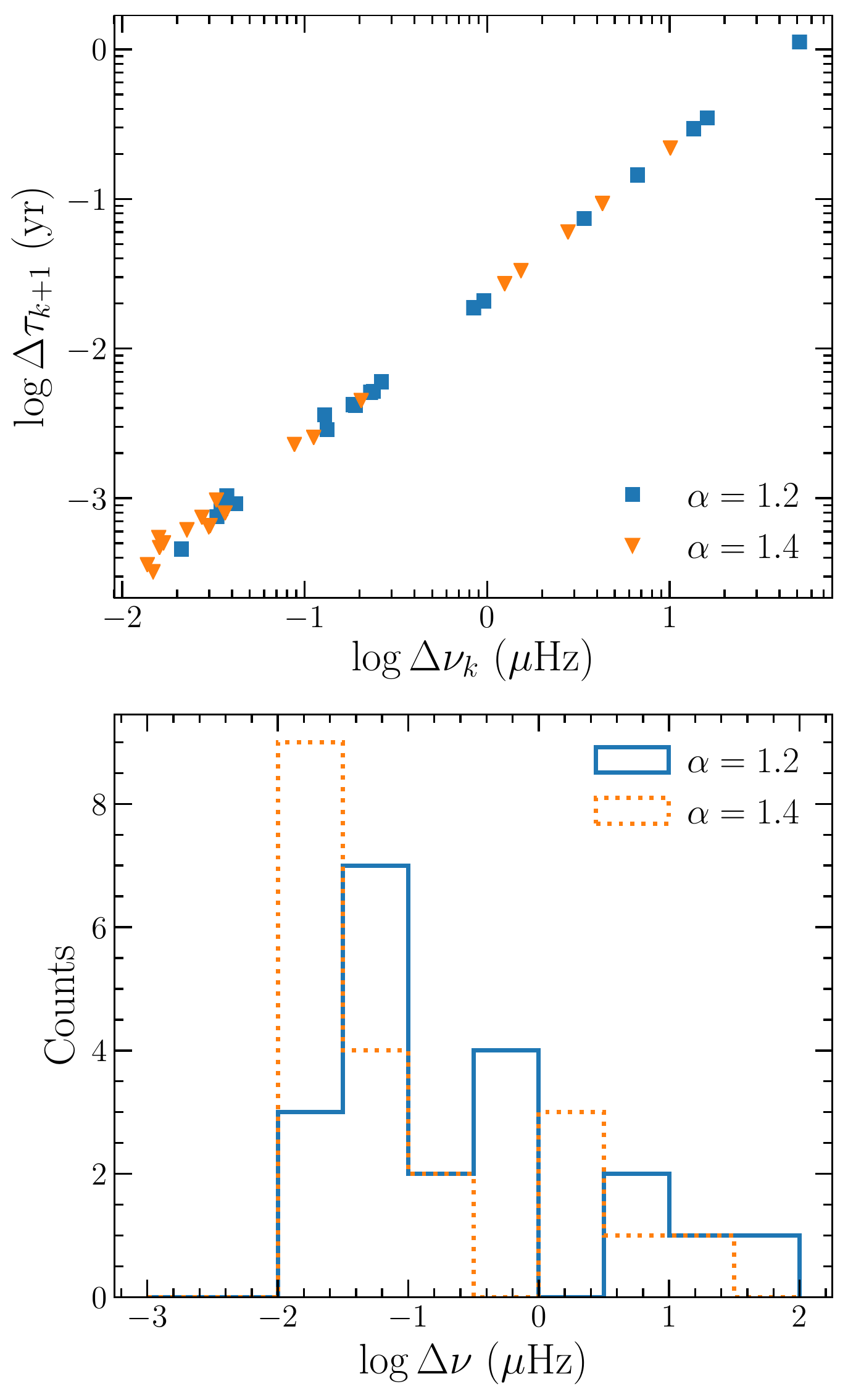}
\caption{Reduced samples of simulated glitches from an assumed parent distribution $dN/d\Delta\nu\propto \Delta\nu^{-\alpha}$ with a perfect correlation $\Delta\tau_{k+1}=C\Delta\nu_k$, with $C=0.21\, \mathrm{yr\, \mu Hz^{-1}}$.
Top: Resulting correlation between $\Delta \nu_k$ and $\Delta \uptau_{k+1}$. 
%in log scale.
Bottom: The corresponding distributions of $\log \Delta \nu$ for the reduced samples of glitches. For both panels, each color (and point marker) represents a typical realization in the simulations, for different power-law exponents as shown in the legends.}\label{hyp1}
\end{figure}

After simulating $10^4$ cases, it was found that removing all glitches smaller than a certain value has a minor effect on the correlation. 
Representative realizations are shown in Fig. \ref{hyp1}, where the correlation between $\Delta \nu_k$ and $\Delta \uptau_{k+1}$ is plotted in log-scale to show more clearly the dispersion produced by the removal of the smallest glitches. 
We observe that missing small glitches does not substantially worsen the correlation: more than $90\%$ of the realizations give correlation coefficients $\geq 0.95$ (both Pearson and Spearman).

For $\alpha>1.4$ the distribution becomes narrower, accumulating towards the lower limit. 
Since a large fraction of the simulated glitches have very similar sizes, after removing the 80 smallest glitches the correlation does worsen, and yields correlation coefficients between $0.4$ and $0.9$, which are similar to those exhibited by the real data.
However, in these cases the distributions of glitch sizes differ strongly from those observed for the pulsars in our sample.

From these simulations, we conclude that it is unlikely that the non-detection of all the glitches below a certain detection limit is the explanation for the low observed correlations in pulsars other than PSR J0537$-$6910.

\subsection{Hypothesis II: Two classes of intrinsically different glitches}

The second hypothesis states that pulsars exhibit two classes of glitches: larger events, which follow a linear correlation between $\Delta \nu_k$ and $\Delta \uptau_{k+1}$; and smaller events, for which these variables are uncorrelated.
We allow the point of separation between large and small glitches to be different for each pulsar.

\begin{figure*}
\centering
\includegraphics[width=180mm]{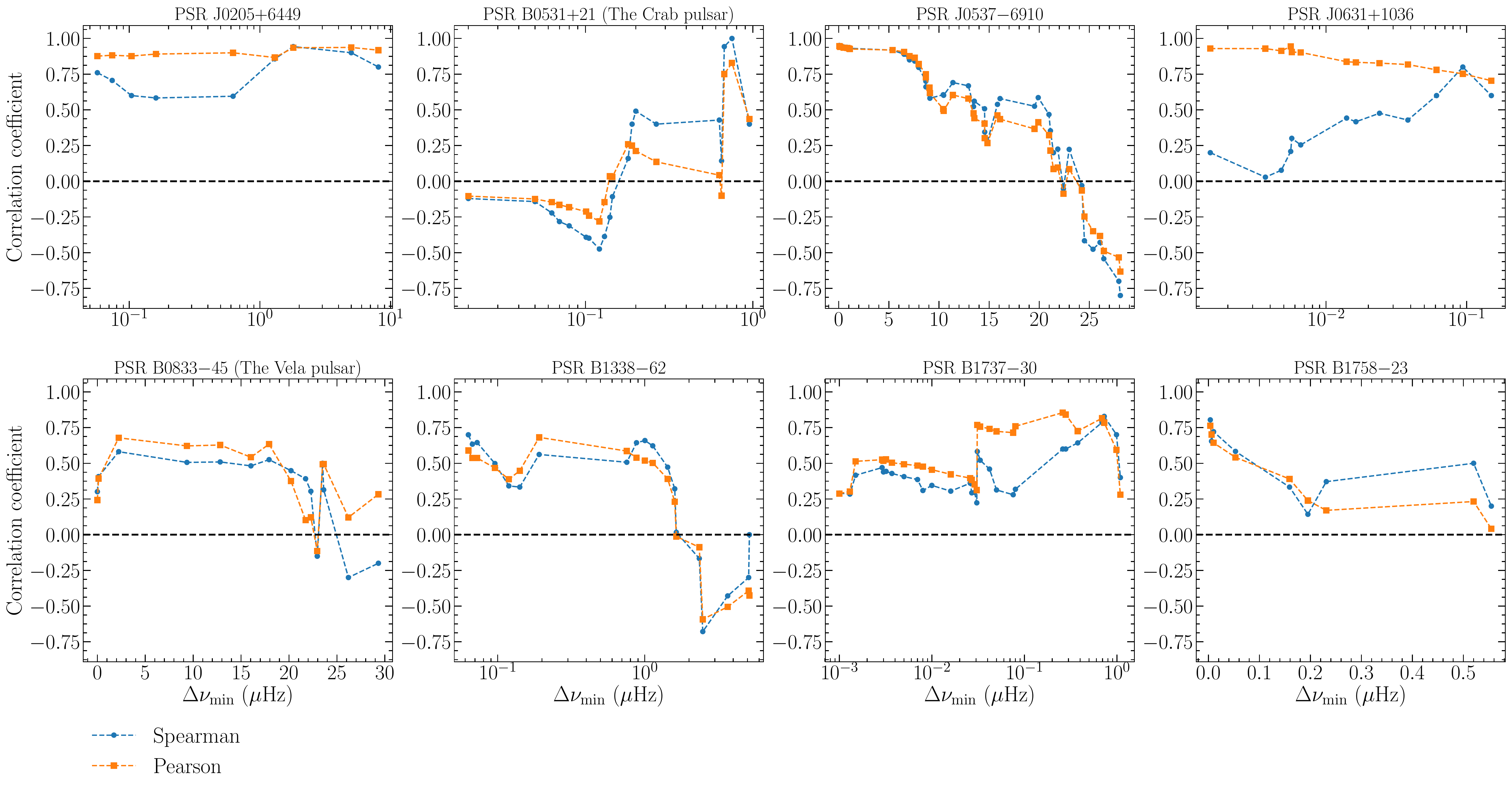}
\caption{Pearson (orange squares) and Spearman (blue dots) correlation coefficients for glitches larger or equal than $\Delta \nu_\textrm{min}$. Each panel represents a pulsar in our sample. For each pulsar, the last point in the plot was calculated with its five largest glitches. Note that some pulsars are shown in log-scale for a better visualization.} 
\label{r_df0_min}
\end{figure*}

To visualize whether this hypothesis works, correlation coefficients (for the same pair of variables, $\Delta \nu_k$ and $\Delta \uptau_{k+1}$) were calculated for sub-sets of glitches of the original sample. 
The sub-sets are defined as all glitches with sizes larger or equal to a given $\Delta\nu_\textrm{min}$. 
Correlation coefficients as a function of $\Delta\nu_\textrm{min}$ are plotted in Fig. \ref{r_df0_min} for each pulsar.
Visual inspection of the plots immediately tells us that by removing small glitches no pulsar reaches the level of correlation observed for PSR J0537$-$6910, for both correlation tests.

In the following we explore the curves in Fig. \ref{r_df0_min} in some more detail.
For that purpose, Monte Carlo simulations of pulsars with correlated and uncorrelated glitches were performed.
Since the underlying glitch size distributions of the pulsars in the sample are unknown, we use the measured values of a given pulsar.
%Hence a set of simulations is related to a particular pulsar.
The following is the procedure for one realization:

\begin{enumerate}
\item The glitches larger than a certain value $\Delta\nu^{\star}$ are chosen in random order and assigned epochs according to their size.
The first one is set at an arbitrary epoch and the epochs of the following ones are assigned according to 
\begin{equation}
\Delta\uptau_{k+1}=\Delta\nu_{k}\cdot 10^{x}\, ,
\label{eq_hyp2}
\end{equation}
where $x$ is drawn from a Gaussian distribution centred at $\bar{x}=\log(C)$ and with a standard deviation equal to $\sigma_{\bar{x}}$. 
The latter allows us to introduce a dispersion in the correlation of the simulated glitches. 
The distribution of $\log(\Delta\uptau_{k+1}/\Delta\nu_k)$ for all glitches with $\Delta\nu>5\,\mu$Hz in PSR J0537$-$6910 can be well modelled by a Gaussian distribution with standard deviation $\sigma_{0537}=0.085$ (in logarithmic scale, if $\Delta\uptau_{k+1}$ is measured in days and $\Delta\nu_k$ is measured in $\mu$Hz). 
In the simulations, $\sigma_{\bar{x}}$ was set either to zero (i.e. $x=\log(C)$, perfect correlation) or to multiples of $\sigma_{0537}$. \\

\item The glitches smaller than $\Delta\nu^{\star}$ are distributed randomly over the time span between the first and the last correlated glitches. 
The resulting waiting times of all, correlated and uncorrelated glitches are then multiplied by a factor that ensures that their sum equals the time in between the first and the last observed glitches. \\

\item Steps 1 and 2 were repeated $10^4$ times for each considered value of $\Delta \nu^{\star}$. 

\end{enumerate}

\begin{figure*}
\centering
\includegraphics[width=18cm]{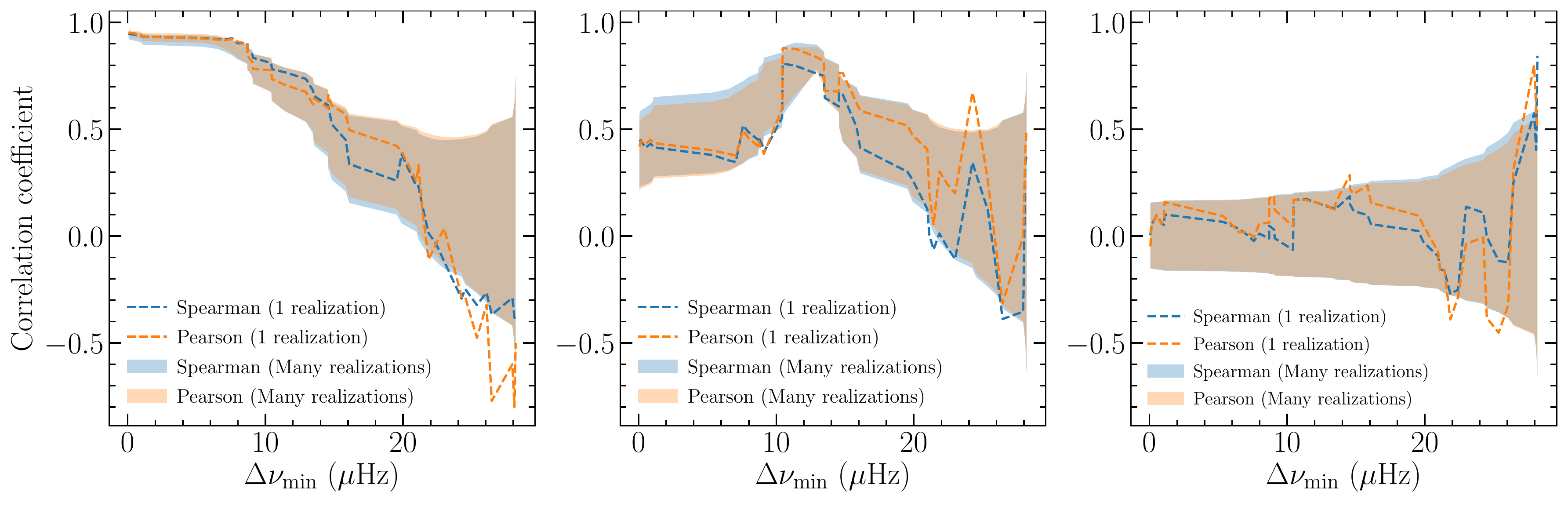}
\caption{Correlation coefficients $r_p$ (orange) and $r_s$ (blue) versus $\Delta\nu_\textrm{min}$ for simulated glitches under hypothesis II, and for three $\Delta\nu^{\star}$ cases: left, when all glitches are correlated ($\Delta\nu^{\star}\sim0$); middle, about half of them are correlated ($\Delta\nu^{\star}=12.39\,\mu$Hz); right, none of them is correlated ($\Delta\nu^{\star}=40\,\mu$Hz).
Shaded regions represent the values of the $70\%$ closer to the median of all realizations.
The dashed lines show particular realizations.
These simulations used the glitch sizes of PSR J0537$-$6910 and $\sigma_{\bar{x}}=\sigma_{0537}$.
In all cases the last points in the plots were calculated using the five largest glitches.
}
\label{examples}
\end{figure*}

The plots in Fig. \ref{examples} show the results of simulations using the glitch sizes of PSR J0537$-$6910 and $\sigma_{\bar{x}}=\sigma_{0537}$ for three values of $\Delta \nu^{\star}$.
The results are shown via curves of $r$ versus $\Delta\nu_\textrm{min}$, to compare with Fig. \ref{r_df0_min}.
The shaded areas represent the $70\%$ of the correlation coefficients closer to the median of all realizations.
We visually inspected the distributions of $r_p$ and $r_s$ for all possible $\Delta\nu_\textrm{min}$  values, and for many $\Delta \nu^{\star}$ cases.
It was verified that the median is sufficiently close to the maximum of the distribution in most cases.
Though, this tends to fail for the largest $\Delta\nu_\textrm{min}$ values, where the $r_p$ and $r_s$ distributions are rather flat.
But this is irrelevant because any conclusion pointing to a case in which only a few glitches are correlated (large $\Delta\nu_\textrm{min}$) would have little statistical value, regardless of the above.
Thus we are confident that the shaded areas effectively cover the most possible outcomes of series of glitches under the assumptions considered.

We now use the plots in Fig. \ref{examples} to understand the curves of the correlation coefficients as functions of $\Delta\nu_\textrm{min}$ in Fig. \ref{r_df0_min}, in the frame of hypothesis~II:

\begin{itemize}

\item[(a)] If all glitches were correlated, which is the case shown in the leftmost plot in Fig. \ref{examples}, the correlation coefficients would decrease gradually as $\Delta\nu_\textrm{min}$ increases. 
This is because a progressive reduction of the sample, starting from the smallest events (i.e. increasing the remaining waiting times by small random amounts), will gradually kill the correlation.
Note that the correlation coefficients of the simulated glitches start at values just below $1.0$ for the smallest $\Delta\nu_\textrm{min}$, just like the observations of PSR J0537$-$6910.
This is because $\sigma_{\bar{x}}=\sigma_{0537}$ in those simulations.
Only for $\sigma_{\bar{x}}=0$ the simulations would start at correlation coefficients equal to $1.0$.
% and would decrease slower than the real data as $\Delta\nu_\textrm{min}$ increases.

\item[(b)] If only glitches above a certain size $\Delta \nu^{\star}$ were correlated, the correlation coefficients would improve as small glitches are eliminated, and the remaining sub-set approaches the one in which all glitches are correlated (as in the middle plot of Fig. \ref{examples}).
One would expect a maximum correlation for $\Delta\nu_\textrm{min}\sim\Delta \nu^{\star}$, and a gradual decrease as $\Delta\nu_\textrm{min}$ increases beyond $\Delta \nu^{\star}$.

\item[(c)] If there were no correlated glitches, we should expect a rather flat curve of low correlation coefficients oscillating around zero (rightmost plot in Fig. \ref{examples}).

\end{itemize}

The behaviours just described correspond to the general trends exhibited by the shaded areas in Fig. \ref{examples}, which evolve smoothly with $\Delta\nu_\textrm{min}$. 
However, particular realizations show abrupt variations, of both signs, just as the observations do in Fig. \ref{r_df0_min}.

Clearly, PSR J0537$-$6910 is best represented by case (a).
Indeed, both correlation coefficients for this pulsar are maximum (and very similar) when all glitches are included and they decrease gradually as the smallest glitches are removed (Fig. \ref{r_df0_min}).
Nonetheless, we note that $r_p$ stays above $0.9$ (and $p_p<3\times10^{-12}$) for $\Delta\nu_\textrm{min}\leq7\,\mu$Hz, hence it is possible that the smallest glitches are not correlated. 
Another indication for this possibility is that the six glitches below $5\,\mu$Hz fall to the right of the  distribution of $\log(\Delta\uptau_{k+1}/\Delta\nu_k)$ for all glitches, and the width of the distribution is reduced considerably (from more than 2 decades to a half decade) when they are removed.
In other words, the straight line that best fits the ($\Delta\uptau_{k+1}$, $\Delta\nu_k$) points passes closer to the origin \citep[a more physically motivated situation,][]{aeka18}, and the data exhibit a smaller dispersion around this line, when the smallest glitches are not included.

The pulsars B1338$-$62, and B1758$-$23 may in principle also correspond to case (a).
As mentioned at the beginning of section \ref{s2}, they present mildly significant correlations when all their glitches are considered, and both their $r_p$ and $r_s$ curves in Fig. \ref{r_df0_min} decrease as $\Delta\nu_\textrm{min}$ increases.
By performing simulations with $\Delta \nu^{\star}=0$, and for different values of $\sigma_{\bar{x}}$, we find that the correlation coefficients of PSR B1758$-$23 are within the range of $70\%$ of the possible outcomes if $\sigma_{\bar{x}}$ is set to 5-6 times $\sigma_{0537}$.

For PSR B1338$-$62 the situation is less clear because the amplitudes of the variations of both $r_p$ and $r_s$ for $\Delta\nu_\textrm{min}<1\,\mu$Hz are rather high.
One possible interpretation is that all glitches are correlated and the variations are due to the correlation not being perfect (i.e. $\sigma_{\bar{x}}\neq0$).
We find that only for $\sigma_{\bar{x}}\geq10\times\sigma_{0537}$ the simulations can reproduce such behaviour and the observed values. 
Another possibility is that $\Delta \nu^{\star}\sim0.2\,\mu$Hz, which could explain the local maxima of $r_p$ and $r_s$ around that value.
The maxima and subsequent values can indeed be reproduced with lower levels of noise, $\sigma_{\bar{x}}=5\times\sigma_{0537}$. 
But for smaller values of $\Delta\nu_\textrm{min}$ most realizations ($>70\%$) give correlation coefficients below $0.5$, thus they fail at reproducing the observed $0.6$-$0.7$ at $\Delta\nu_\textrm{min}=0$.

It is clear that Hypothesis II does not apply to this pulsar directly, and that the observations are not consistent with a set of uncorrelated glitches either.
Based on the lack of glitches with sizes equal or less than $0.1\,\mu$Hz after MJD $\sim$ 50400 (Fig. \ref{fig2}), we speculate that the sample might be incomplete for glitches smaller than this size after this date\footnote{This would be a more extreme case than those considered for the Hypothesis I because $0.1\,\mu$Hz is a rather high limit.}.

The pulsars J0205$+$6449 and J0631$+$1036 also exhibit significant Pearson correlations when all their glitches are considered.
However, their $r_s$ curves tend to increase with $\Delta\nu_\textrm{min}$ rather to decrease.
As mentioned before, the Pearson test can be affected by outliers, hence the behaviour we see for $r_p$ is likely due to the very broad size and waiting times distributions and the low numbers of events towards the high ends of the distributions, which produce outlier points for both pulsars (Fig.\ref{fig6}).
It is therefore difficult to conclude anything for PSR J0631+1036. 
Moreover, the observed behaviour is very hard to reproduce by the simulations, even for high levels of noise (we tried up to $\sigma_{\bar{x}}=12\times\sigma_{0537}$).
Perhaps its largest glitches ($\Delta\nu\geq0.1\,\mu$Hz) are indeed correlated, but the statistics are too low to conclude anything.

For J0205, however, the Spearman coefficients $r_s$ are rather high ($>0.55$ for all $\Delta\nu_\textrm{min}$) and both coefficients become similar and even higher for $\Delta\nu_\textrm{min}>1\,\mu$Hz. 
It is possible that glitches above this size are correlated in this pulsar.
We find that the observed $r_p$ and $r_s$, and their evolution with $\Delta\nu_\textrm{min}$, are within the $70\%$ of simulations with $\Delta \nu^{\star}=1.3\,\mu$Hz and for $\sigma_{\bar{x}} = 2\times\sigma_{0537}$. 
We note, however, that in this case the correlation coefficients observed for  $\Delta\nu_\textrm{min}\leq0.1\,\mu$Hz are higher than the vast majority of the realizations.
Perhaps the small glitches are also correlated and follow their own relation, though we did not simulate such scenario.
We conclude that the Hypothesis II does not fully explain this pulsar, although the 8 glitches above $1\,\mu$Hz appear to be well correlated indeed.

\begin{figure*}
\centering
\includegraphics[width=18cm]{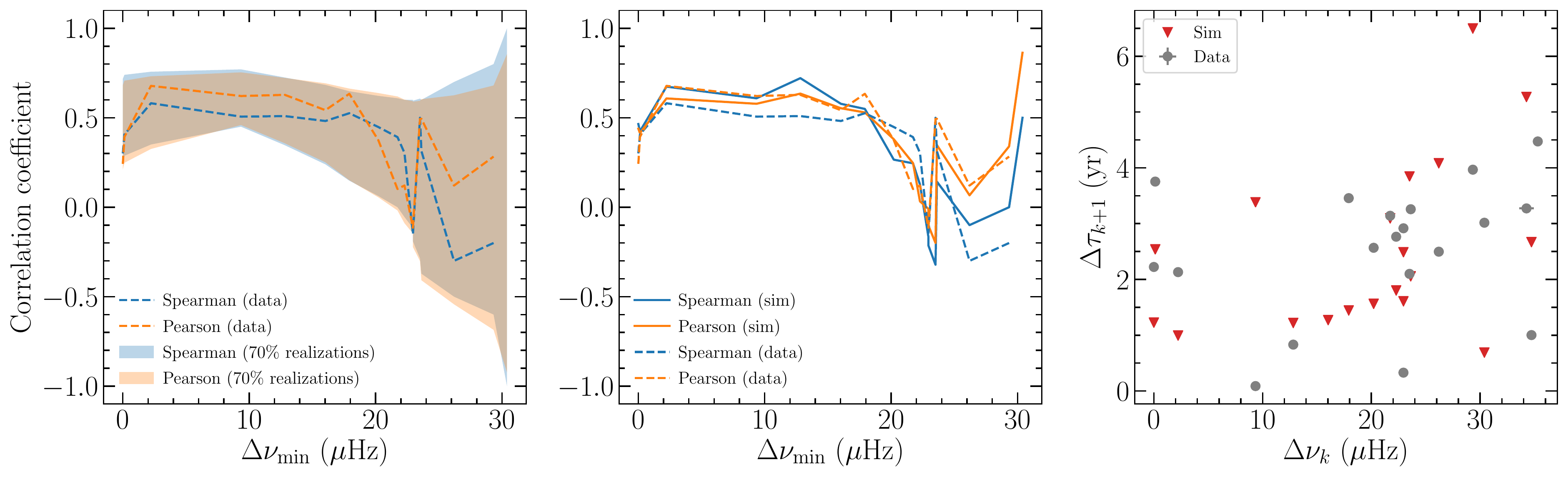}
\caption{Observations and simulations of the Vela pulsar.
Left: Shaded regions indicate the values obtained by the $70\%$ closer to the median of all realizations. 
The observations are overlaid using dashed lines.
Centre: comparison of observations (dashed) and one particular realization.
Right: $\Delta\uptau_{k+1}$ versus $\Delta\nu_k$ for the same realization (red triangles) and for the observations (grey dots).
Orange represents $r_p$ values and blue represents $r_s$ values in all panels.
The simulations were performed using $\sigma_{\bar{x}}=2\times\sigma_{0537}$ and $\Delta\nu^{\star}=9.35\,\mu$Hz.}
\label{vela_h2}
\end{figure*}

The Vela pulsar is the only pulsar in the sample that seems well represented by case (b).
The highest $r_p=0.68$ has a probability $p_p=0.003$ and is obtained for $\Delta\nu_\textrm{min}\sim2\,\mu$Hz.
Both $r_p$ and $r_s$ decline monotonically for larger $\Delta\nu_\textrm{min}$ values. 
This behaviour suggests that glitches of sizes above $\sim2\,\mu$Hz might indeed be correlated, but the correlation is somewhat noisy.
The observed correlation coefficients fall within the middle 70$\%$ of the realizations if $\sigma_{\bar{x}}=2\times\sigma_{0537}$ and for $\Delta\nu^{\star}=2$-$10\,\mu$Hz.
The case $\Delta\nu^{\star}=9.35\,\mu$Hz is presented in Fig. \ref{vela_h2}.
We prefer this case because simulations for $\Delta\nu^{\star}=2\,\mu$Hz tend to fail at reproducing the low correlation coefficients ($\leq0.4$) observed for the smallest $\Delta\nu_\textrm{min}$.

Finally, the cases of PSRs B0531+21 (the Crab) and B1737$-$30 are rather inconclusive.
The Crab pulsar is perhaps the pulsar for which case (c) applies the best. 
Both correlation coefficients are negative or positive, and in both cases stay at relatively low absolute values, which leads to the conclusion that there are no correlated glitches in the Crab pulsar.
We note that the high $r_p$ and $r_s$ values observed for $\Delta\nu_\textrm{min}\sim0.6\,\mu$Hz are obtained with the 5-6 largest events and that a linear fit to their $\Delta \nu_k - \Delta\uptau_{k+1}$ does not pass close to the origin.

The case of B1737$-$30 is more complex.
The observations show two $\Delta\nu_\textrm{min}$ values, $0.0015$ and $0.03\,\mu$Hz, after which the correlation coefficients decrease with the removal of more small glitches (Fig. \ref{r_df0_min}). 
This behaviour is hard to reproduce under Hypothesis II, unless the dispersion of the correlation is increased considerably, to $10\times\sigma_{0537}$ or more.
We conclude that Hypothesis II does not apply to this pulsar directly and that there is some extra complexity, as the data are also inconsistent with a set of purely uncorrelated glitches.

Surprisingly, even though no pulsar complies perfectly with Hypothesis II, and the only way in some cases is to increase the dispersion of the correlation ($\sigma_{\bar{x}}\gg\sigma_{0537}$), there is no pulsar in the sample that is well represented by case (c) (only the Crab, to some extent).

Therefore, the sizes of at least some glitches must be positively correlated with the times to the next glitch in the available datasets.
The question is why this correlation is much stronger in PSR J0537$-$6910 than in all other pulsars of our sample.
Could this be an effect of its particularly high spin-down rate? Or the fact that most of its glitches are large?
It could be that the correlations are indeed there, as stated in Hypothesis II, but for some reason exhibit high $\sigma_{\bar{x}}$ values. 
Maybe the fact that the glitches in PSR J0537$-$6910 occur so frequently ensures that the relationship stays pure.
But it could also be that reality was more complex.
For instance, it could be that both small and large glitches were correlated, but each of them followed a different law.

\section{Other correlations} 
\label{s3}

We looked for other correlations between the glitch sizes and the times between them.
Specifically, we tried  $\Delta \nu_k$ vs $\Delta \uptau_{k}$ (size of the glitch versus the time since the preceding glitch), and $\Delta \nu_k$ vs $\Delta \nu_{k-1}$ (size of the glitch versus the size of the previous glitch).
No pulsar shows a significant correlation between these quantities (Table \ref{others_correlations}).

\begin{table*}
\caption{Correlation coefficients for the pairs of variables $(\Delta \nu_k,\,\Delta \uptau_{k})$, and $(\Delta \nu_k,\, \Delta \nu_{k-1})$.} \label{others_correlations}
\small % reduce font size by 1pt
     \begin{subtable}{0.47\textwidth}
     \begin{tabular*}{\linewidth}{@{}l 
        @{\extracolsep{\fill}} SS
        S[table-format=2.2(2)]
        S[table-format=2.2(2)]@{}}
          \toprule
          \phantom{Var.} &  
          \multicolumn{4}{c}{$\Delta \nu_k$ vs $\Delta \uptau_{k}$}\\
          \cmidrule{1-5}
          {PSR Name}&  {$r_p$} & {$p_p$} & {$r_s$} & {$p_s$}\\
          \midrule
          J0205$+$6449\hspace{0.5cm}  & 0.16 & 0.60 & 0.44 & 0.15\\
          B0531$+$21   & -0.02 & 0.90 & 0.40 & 0.05\\
          J0537$-$6910   & -0.08 & 0.60  & -0.12 & 0.41 \\[1ex]
          J0631$+$1036 & -0.10 & 0.68 & -0.18 & 0.49 \\ % 1e7
          B0833$-$45 &  0.55 & 0.01 & 0.27 & 0.24 \\
          B1338$-$62 &  -0.30 & 0.16 & -0.18 & 0.41 \\[1ex]
          B1737$-$30 &  -0.02 & 0.89 & -0.10 & 0.56 \\ % 1e5
          B1758$-$23 &  -0.02 & 0.94 & -0.04 & 0.89 \\
          \bottomrule
     \end{tabular*}%
     %\caption{}
     \end{subtable}%
     \hspace*{\fill}%
     \begin{subtable}{0.47\textwidth}
     \begin{tabular*}{\linewidth}{@{}l 
        @{\extracolsep{\fill}} SS
        S[table-format=2.2(2)]
        S[table-format=2.2(2)]@{}}
          \toprule
          \phantom{Var.}
          &  \multicolumn{4}{c}{$\Delta \nu_k$ vs $\Delta \nu_{k-1}$}\\
          \cmidrule{1-5}
          {PSR Name}&  {$r_p$} & {$p_p$} & {$r_s$} & {$p_s$}\\
          \midrule
          J0205$+$6449\hspace{0.5cm}  & -0.06 & 0.83 & 0.25 & 0.42\\
          B0531$+$21  & -0.10 & 0.61 & -0.15 & 0.47\\
          J0537$-$6910  & -0.13 & 0.38 & -0.16  & 0.29\\[1ex]
          J0631$+$1036 & -0.12 & 0.65 & 0.32 & 0.21 \\ % 1e7
          B0833$-$45 & -0.08 & 0.71 & -0.12 & 0.59\\
          B1338$-$62 & -0.33 & 0.13 & -0.13 & 0.55\\[1ex]
          B1737$-$30 & -0.11 & 0.50 & 0.03 & 0.85\\ % 1e5
          B1758$-$23 & -0.02 & 0.92 & -0.04 & 0.89\\
          \bottomrule
     \end{tabular*}%
     %\caption{}
     \end{subtable}
\tablefoot{The first column contains the names of the pulsars considered in the sample. $r_{n}$ and $p_{n}$ correspond to a correlation coefficient and its $p$-value, respectively. The sub-index $n = p$ denotes the Pearson correlation, and  $n=s$ denotes the Spearman correlation.}
\end{table*}

Nearly all pulsars in our sample show negative correlation coefficients (both, Pearson and Spearman) for $\Delta \nu_k$ vs $\Delta \uptau_{k}$. 
The only exceptions are the Vela pulsar and PSR J0205$+$6449 (see Table \ref{others_correlations}). Our results are in general agreement with \cite{mhf18}, %although they found low and positive correlation coefficients for the Crab, Vela, and PSR J0537$-$6910. This discrepancy could be due to small differences in the samples. However, both results are consistent with 
who also found a lack of correlation between $\Delta \nu_k$ vs $\Delta \uptau_{k}$ for individual pulsars.

For $\Delta \nu_k$ vs $\Delta \nu_{k-1}$, in most cases the correlation coefficients are close to zero and the $p$-values are larger than $0.2$, i.e., no individual pulsar shows a significant correlation. However, the results could still be meaningful for the sample as a whole because all the pulsars have negative correlation coefficients, except for the Spearman coefficients for PSRs J0631$+$1036 and B1737$-$30). 
The probability of getting all Pearson's correlations coefficients of the same sign just by chance, regardless of whether the sign is positive or negative, is $2\times p_{\mathrm{binom}}(8|8) = 0.007$. This could establish an interesting constraint on the glitch mechanism: Smaller glitches are somewhat more likely to be followed by larger ones, and vice-versa.
However, this statement has to be confirmed with more data in the future.

\section{Discussion} 
\label{disc}

\begin{figure}
\centering
\includegraphics[width=9cm]{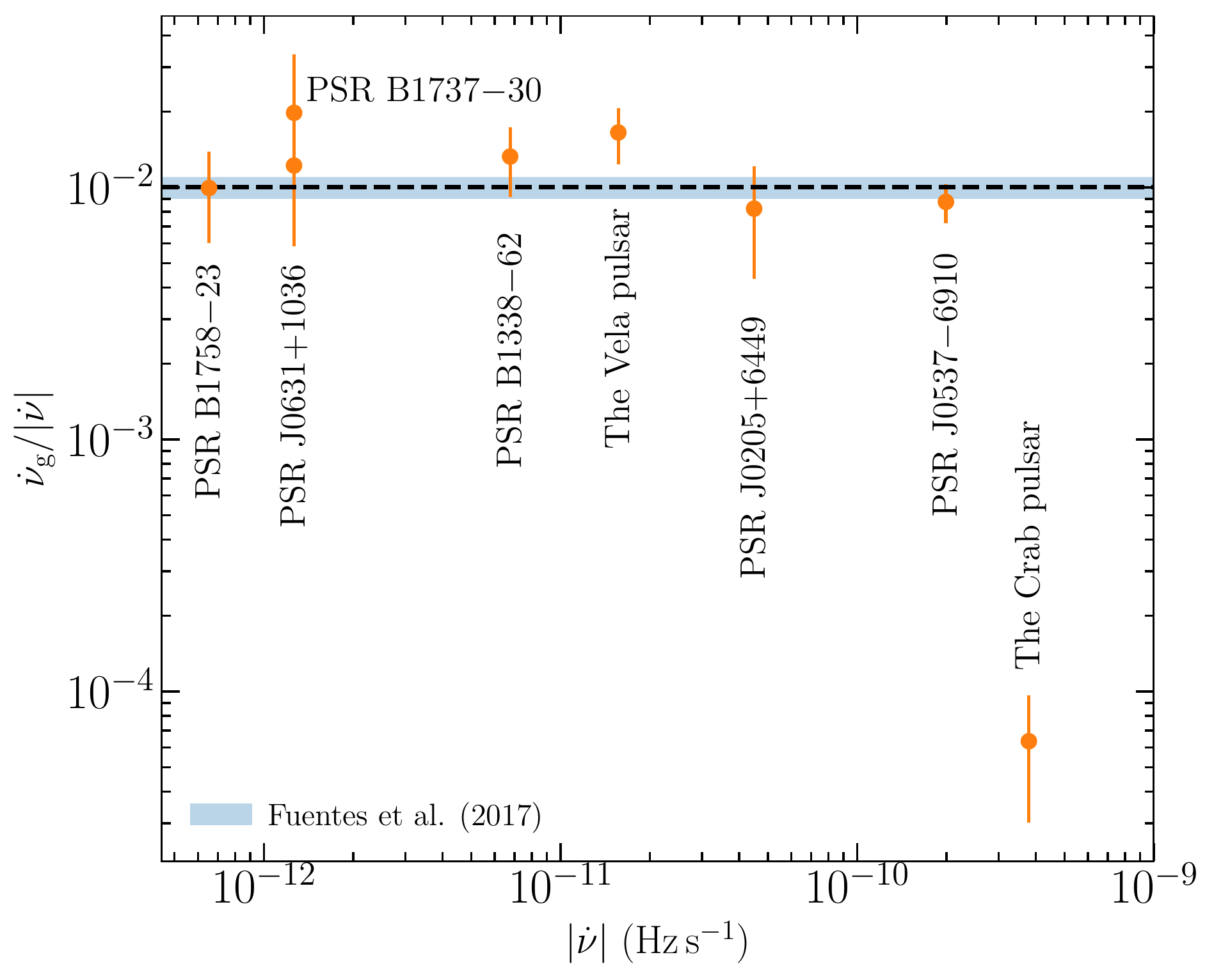}
\caption{$\dot{\nu}_g/|\dot{\nu}|$ versus $|\dot{\nu}|$ for pulsars in our sample.
The dashed-line with the blue region correspond to the constant ratio $\dot{\nu}_g/|\dot{\nu}| = 0.010 \pm
0.001$, determined by \citet{fer+17}.
The error bars were calculated as described in the latter paper.}
\label{fig_discussion}
\end{figure}

\citet{fer+17} found that all pulsars (with the strong exception of the Crab pulsar and PSR B0540$-$69) %\borrar{see also \citealt{elsk11}}) 
are consistent with a constant ratio between the glitch activity, $\dot{\nu}_{\rm g}$, and the spin-down rate, %measured the relation 
$\dot\nu_{\rm{g}}/|\dot\nu| = 0.010 \pm 0.001$, i.e., $\approx 1\%$ of their spin-down is recovered by the glitches. This fraction has been interpreted as the fraction of the moment of inertia in a superfluid component that transfers its angular momentum to the rest of the star in the glitches  \citep{lel99,aghe12}. %where $\dot{\nu}_{\rm g}$ is the glitch activity, and remarked that this relation is determined by the large glitches.
%\footnote{As noted by \cite{elsk11} and \cite{mhf18}, and confirmed by our results, the Crab pulsar and PSR B0540$-$69 exhibit an exceptional behaviour that is off this trend. Consequently, we do not try to understand their behaviour in line with the rest of the pulsars.} \citet{fer+17} also showed that it is not possible to rule out the possibility that all pulsars satisfy such a relationship.
\citet{fer+17} used the observed bimodal distribution of glitch sizes to distinguish between large and small glitches, with the boundary at $\Delta\nu=10\, \mathrm{\mu Hz}$, and argued that the constant ratio is determined by the large glitches, whose rate, $\dot N_\ell$ is also proportional to $|\dot\nu|$. In %such a 
this scenario, the %reason why many low-$|\dot{\nu}|$ pulsars show 
much lower (sometimes null) glitch activities measured in many low-$|\dot{\nu}|$ pulsars are due to 
%is because the 
their observation time spans %are 
not being long enough to include any large glitches (or any glitch at all). %in those pulsars.
Interestingly, the pulsars in our sample (except the Crab) are quite consistent with the constant ratio (Fig. \ref{fig_discussion}), even those, like PSRs B1338$-$62, B1737$-$30, and B1758$-23$, which do not have any large glitches contributing to their activities.
%For example, PSR B1737$-$30 has only small glitches and its glitch activity satisfies $\dot\nu_{\rm{g}}/|\dot\nu| =0.013$, in good agreement with the above relationship.
%Therefore, it seems that the \ar{spin-up} by glitches in these pulsars remains at $\sim 1$\% of $|\dot{\nu}|$ regardless of the glitch sizes and how they are distributed. 
%This suggests that both small and large glitches draw from a common reservoir of angular momentum.
%In fact, this $1$\% is interpreted as the fraction of the total moment of inertia that participates in the generation of glitches, i.e. the relative size of the glitch-driving superfluid \citep{lel99,aghe12}.

%On the other hand, pulsars that are slowing down faster generally exhibit more glitches during a given time than those evolving \ar{more slowly.} %slower.
%By considering more than 600 pulsars and grouping them according to their spin-down rates, \citet{elsk11} found a relationship in which the glitch rate $\dot{N}_g$ is roughly proportional to $|\dot{\nu}|^{1/2}$.
%\citet{fer+17} made a distinction between small and large glitches (with the \ar{boundary} at $10\,\mu$Hz) and established that the rate of large glitches satisfies  $\dot{N}_\ell\propto|\dot{\nu}|$.
%Thus the rate %for 
%\ar{of} large glitches decays much faster, as $|\dot{\nu}|$ gets smaller, than the rate %for 
%\ar{of} small glitches.
%Only f

On the other hand, pulsars with higher spin-down rates also have a larger fraction of large glitches. At the highest spin-down rates ($|\dot{\nu}|\geq 10^{-11}$\,Hz\,s$^{-1}$), the production of large glitches becomes comparable and sometimes higher than the production of small glitches, again with the notorious exception of the Crab and PSR B0540$-$69.
This trend is also followed by the pulsars in our sample: all large glitches (but one in PSR J0631+1036), are concentrated in PSRs J0205$+$6449, J0537$-$6910, and the Vela pulsar, which are (together with the Crab) the ones with largest $|\dot{\nu}|$ values (see Fig. \ref{fig1} and \ref{fig_discussion}).

Thus, it seems to be the case that both large and small glitches draw from the same angular momentum reservoir (for all but the very young, Crab-like pulsars), but have different trigger mechanisms, the large ones being produced once a critical state %threshold angular velocity difference 
is reached, whereas small ones occur in a more random fashion. 
For reasons still to be understood, the glitch activity of relatively younger, high $|\dot\nu|$, Vela-like pulsars is dominated by large glitches, whereas for smaller $|\dot \nu|$ the large glitches become less frequent, both in absolute terms and relative to the small ones \citep{wmp+00,elsk11}. 
%, and as their spin-down rate decreases and they evolve towards 
%\ar{Thus}, it remains unclear what determines the particular glitching style of a pulsar; whether it will glitch like Vela or like PSR B1737$-$30, for instance. 
%One possibility could be age, or evolution of the glitch mechanism due to the change of relevant  properties.
%This would be in line with 
In this context, it is interesting to note that recent long-term braking index measurements %that 
indicate that Vela-like pulsars move  towards the region where PSRs J0631+1036, B1737$-$30, and B1758$-$23 are located on the $P$--$\dot{P}$ diagram \citep[][]{els17}.

\section{Summary and Conclusions} 
\label{conc}

We studied the individual glitching behaviour of the eight pulsars that today have at least ten detected glitches.
Our main conclusions are the following:

\begin{enumerate}

\item 
We confirm the previous result by \cite{mpw08} and \cite{hmd18} that, for Vela and PSR J0537$-$6910, the distributions of both their glitch sizes and waiting times are best fitted by Gaussians, indicating well-defined scales for both variables. For all other pulsars studied, the waiting time distribution is best fitted by an exponential (as would be expected for mutually uncorrelated events), but they have a variety of best-fitting size distributions: a power law for PSR J0205+6449, J0631+1036, and B1737$-$30, a log-normal for the Crab and PSR B1338$-$62, and an exponential for PSR B1758$-$23. %\rafa{These results could be explained within the stochastic model of \cite{cm19}, who showed that a variety of size probability density function shapes apart from power-law and Gaussian emerge from a state-dependent Poisson model, which is controlled by a parameter related to the spin-down rate of the pulsar.}
%are  The \ar{size} distributions of the eight pulsars studied in this paper cannot be described by a single type of distribution.
%Instead, we find that different pulsars are best described by four different distribution functions.
%The waiting time distributions show only two groups: exponential laws (six cases) and Gaussian functions (two cases).
%The latter pulsars are Vela and J0537$-$6910, which also exhibit Gaussian size distributions centred at large sizes ($>10\,\mu$Hz).
%The other six studied pulsars show exponential waiting time distributions, as it would be expected from a process in which glitches were randomly distributed in time.
%However, this is not precisely like this, as we showed how glitch sizes have at least some mild  influence on the time to the next glitch (see below). \\

\item 
All pulsars in our sample, except for the Crab, have positive Spearman and Pearson correlation coefficients for the relation between the size of each glitch, $\Delta\nu_k$, and the waiting time to the following glitch, $\Delta\tau_{k+1}$. For each coefficient, the probability for this happening by chance is $1/16=6.25\%$. 
Both coefficients also stay positive as the small glitches are removed %most correlation coefficients are positive in 
(see Fig. \ref{r_df0_min}).
%\cme{Our simulations also show that no pulsar in the sample  (except for the Crab) is consistent with a complete lack of correlation.
%Thus the size of glitches somehow determines the time to the next glitch. }

%This result is \ar{significant for the sample as a whole, having a probability of 6.25$\%$ of happening by chance.}
%The time to the next glitch is therefore somewhat regulated by the size of the current glitch. \\

\item 
PSR J0537$-$6910 shows by far the strongest correlation between glitch size and waiting time until the following glitch ($r_p=r_s=0.95$, $p$-values $\lesssim 10^{-22}$). 
Another three pulsars, PSRs J0205$+$6449, B1338$-$62, and B1758$-$23, have quite significant correlations ($p$-values $\leq 0.004$ for both coefficients).
%, \borrar{whereas the weakest correlations are found for Vela and the Crab, the latter having $r_p\approx r_s\approx -0.1$.}
%B1758$-$23 shows a significant correlation but not at the high level observed for PSR J0537$-$6910.
%With the help of MC simulations, we also showed that glitches in the pulsars J0205$+$6449 and Vela (B0833$-$45) may be correlated, but only for sizes above $1$ and $10\,\mu$Hz, respectively. It remains unclear why this dependancy is weaker in most pulsars or what contaminates the data. \\

\item
Our first hypothesis to explain the much weaker correlations in all other pulsars compared to PSR J0537$-$6910, namely missing glitches that are too small to be detected, is very unlikely to be correct. Our Monte Carlo simulations show that, for reasonable glitch size distributions, it cannot produce an effect as large as observed.

\item 
Our alternative hypothesis, namely that there are two classes of glitches, large correlated ones and small uncorrelated ones, comes closer to reproducing the observed relations; notably for PSRs J0205$+$6449 and Vela.
The resulting correlations for both pulsars present dispersions that are twice the one observed for PSR J0537$-$6910.
For the other pulsars, the required dispersions to accommodate this hypothesis are much larger. %\borrar{ and their behaviours are harder to accommodate under this hypothesis.}

%but it still requires a larger dispersion in other pulsars compared to PSR J0537$-$6910.

\item
The correlation coefficients between the sizes of two successive glitches, $\Delta\nu_{k-1}$ and $\Delta\nu_k$, as well as between the size of a glitch, $\Delta\nu_k$ and the waiting time since the previous glitch, $\Delta\tau_k$, are generally not significant in individual pulsars, but they are negative for most cases, suggesting some (weaker) relation also among these variables.

\item
Except for the Crab, all pulsars in our sample are consistent with the constant ratio between glitch activity and spin-down rate, $\dot\nu_\mathrm{g}/|\dot\nu|=0.010\pm 0.001$ \citep{fer+17}. This includes cases dominated by large glitches, as well as others with only small glitches. 

\item
The previous results suggest that large and small glitches draw their angular momentum from a common reservoir, although they might be triggered by different mechanisms. Large glitches, which dominate at large $|\dot\nu|$ (except for the Crab and PSR B0540$-$69), might occur once a certain critical state
%threshold angular velocity difference 
is reached, while small glitches, dominating in older pulsars with lower $|\dot\nu|$, occur at essentially random times.

%The observed behaviour might be interpreted in terms of two different triggers for the glitches, one of which  dominate on high-$|\dot{\nu}|$ pulsars and produce large glitches.
%The other one would start operating at lower $|\dot{\nu}|$ values.
%We speculate that pulsars could transition between these triggers as they age and their spin-down rates decrease.

\end{enumerate}

All the above is based on the behaviour of the pulsars with the most detected glitches. 
Even though we have shown before that the activity of all pulsars appears to be consistent with one single trend, these pulsars could still be outliers among the general population. 
Only many more years of monitoring will clarify the universality of these results.

\begin{acknowledgements}
We thank Vanessa Graber and Simon Guichandut for valuable comments on the first draft of this article. We are also grateful to Wilfredo Palma for conversations that guided us at the beginning of this work. We also thank Ben Shaw for information regarding the detection of recent glitches and for keeping the glitch catalog up to date.
This work was supported in Chile by CONICYT, through the projects ALMA31140029, Basal AFB-170002, and FONDECYT/Regular 1171421 and 1150411.
J.R.F. acknowledges	partial support by an NSERC Discovery Grant awarded to A. Cumming at McGill University.
C.M.E. acknowledges support by the Universidad de Santiago de Chile (USACH).
\end{acknowledgements}

\bibliographystyle{aa}
\bibliography{journals,pulsar}

\begin{thebibliography}{32}
\expandafter\ifx\csname natexlab\endcsname\relax\def\natexlab#1{#1}\fi

\bibitem[{{Akaike}(1974)}]{aka74}
{Akaike}, H. 1974, IEEE Transactions on Automatic Control, 19, 716

\bibitem[{{Anderson} \& {Itoh}(1975)}]{ai75}
{Anderson}, P.~W. \& {Itoh}, N. 1975, Nature, 256, 25

\bibitem[{{Andersson} {et~al.}(2012){Andersson}, {Glampedakis}, {Ho}, \&
  {Espinoza}}]{aghe12}
{Andersson}, N., {Glampedakis}, K., {Ho}, W.~C.~G., \& {Espinoza}, C.~M. 2012,
  Phys. Rev. Lett., 109, 241103

\bibitem[{{Antonopoulou} {et~al.}(2018){Antonopoulou}, {Espinoza}, {Kuiper}, \&
  {Andersson}}]{aeka18}
{Antonopoulou}, D., {Espinoza}, C.~M., {Kuiper}, L., \& {Andersson}, N. 2018,
  \mnras, 473, 1644

\bibitem[{{Ashton} {et~al.}(2017){Ashton}, {Prix}, \& {Jones}}]{apj17}
{Ashton}, G., {Prix}, R., \& {Jones}, D.~I. 2017, \prd, 96, 063004

\bibitem[{Buchner(2013)}]{buc13}
Buchner, S. 2013, The Astronomer's Telegram, 5406

\bibitem[{{Carlin} \& {Melatos}(2019)}]{cm19}
{Carlin}, J.~B. \& {Melatos}, A. 2019, \mnras, 483, 4742

\bibitem[{{Downs}(1981)}]{downs81}
{Downs}, G.~S. 1981, \apj, 249, 687

\bibitem[{{Espinoza} {et~al.}(2014){Espinoza}, {Antonopoulou}, {Stappers},
  {Watts}, \& {Lyne}}]{eas+14}
{Espinoza}, C.~M., {Antonopoulou}, D., {Stappers}, B.~W., {Watts}, A., \&
  {Lyne}, A.~G. 2014, MNRAS, 440, 2755

\bibitem[{{Espinoza} {et~al.}(2017){Espinoza}, {Lyne}, \& {Stappers}}]{els17}
{Espinoza}, C.~M., {Lyne}, A.~G., \& {Stappers}, B.~W. 2017, MNRAS, 466, 147

\bibitem[{{Espinoza} {et~al.}(2011){Espinoza}, {Lyne}, {Stappers}, \&
  {Kramer}}]{elsk11}
{Espinoza}, C.~M., {Lyne}, A.~G., {Stappers}, B.~W., \& {Kramer}, M. 2011,
  MNRAS, 414, 1679

\bibitem[{{Ferdman} {et~al.}(2018){Ferdman}, {Archibald}, {Gourgouliatos}, \&
  {Kaspi}}]{fagk18}
{Ferdman}, R.~D., {Archibald}, R.~F., {Gourgouliatos}, K.~N., \& {Kaspi}, V.~M.
  2018, \apj, 852, 123

\bibitem[{{Fuentes} {et~al.}(2017){Fuentes}, {Espinoza}, {Reisenegger}, {Shaw},
  {Stappers}, \& {Lyne}}]{fer+17}
{Fuentes}, J.~R., {Espinoza}, C.~M., {Reisenegger}, A., {et~al.} 2017, \aap,
  608, A131

\bibitem[{{Fulgenzi} {et~al.}(2017){Fulgenzi}, {Melatos}, \& {Hughes}}]{fmh17}
{Fulgenzi}, W., {Melatos}, A., \& {Hughes}, B.~D. 2017, \mnras, 470, 4307

\bibitem[{Hobbs {et~al.}(2004)Hobbs, Lyne, Kramer, Martin, \& Jordan}]{hlk+04}
Hobbs, G., Lyne, A.~G., Kramer, M., Martin, C.~E., \& Jordan, C. 2004, MNRAS,
  353, 1311

\bibitem[{{Howitt} {et~al.}(2018){Howitt}, {Melatos}, \& {Delaigle}}]{hmd18}
{Howitt}, G., {Melatos}, A., \& {Delaigle}, A. 2018, \apj, 867, 60

\bibitem[{{Konar} \& {Arjunwadkar}(2014)}]{ka14b}
{Konar}, S. \& {Arjunwadkar}, M. 2014, in Astronomical Society of India
  Conference Series, Vol.~13, 87--88

\bibitem[{{Link} {et~al.}(1999){Link}, {Epstein}, \& {Lattimer}}]{lel99}
{Link}, B., {Epstein}, R.~I., \& {Lattimer}, J.~M. 1999, Phys. Rev. Lett., 83,
  3362

\bibitem[{{Lyne} {et~al.}(2000){Lyne}, {Shemar}, \& {Graham-Smith}}]{lsg00}
{Lyne}, A.~G., {Shemar}, S.~L., \& {Graham-Smith}, F. 2000, MNRAS, 315, 534

\bibitem[{McCulloch {et~al.}(1987)McCulloch, Klekociuk, Hamilton, \&
  Royle}]{mkhr87}
McCulloch, P.~M., Klekociuk, A.~R., Hamilton, P.~A., \& Royle, G. W.~R. 1987,
  Aust. J. Phys., 40, 725

\bibitem[{McKenna \& Lyne(1990)}]{ml90}
McKenna, J. \& Lyne, A.~G. 1990, Nature, 343, 349

\bibitem[{{Melatos} {et~al.}(2018){Melatos}, {Howitt}, \& {Fulgenzi}}]{mhf18}
{Melatos}, A., {Howitt}, G., \& {Fulgenzi}, W. 2018, \apj, 863, 196

\bibitem[{{Melatos} {et~al.}(2008){Melatos}, {Peralta}, \& {Wyithe}}]{mpw08}
{Melatos}, A., {Peralta}, C., \& {Wyithe}, J.~S.~B. 2008, ApJ, 672, 1103

\bibitem[{{Middleditch} {et~al.}(2006){Middleditch}, {Marshall}, {Wang},
  {Gotthelf}, \& {Zhang}}]{mmw+06}
{Middleditch}, J., {Marshall}, F.~E., {Wang}, Q.~D., {Gotthelf}, E.~V., \&
  {Zhang}, W. 2006, ApJ, 652, 1531

\bibitem[{Radhakrishnan \& Manchester(1969)}]{rm69}
Radhakrishnan, V. \& Manchester, R.~N. 1969, Nature, 222, 228

\bibitem[{Reichley \& Downs(1969)}]{rd69}
Reichley, P.~E. \& Downs, G.~S. 1969, Nature, 222, 229

\bibitem[{Shemar \& Lyne(1996)}]{sl96}
Shemar, S.~L. \& Lyne, A.~G. 1996, MNRAS, 282, 677

\bibitem[{Wang {et~al.}(2012)Wang, Wang, Tong, \& Yuan}]{wwty12}
Wang, J., Wang, N., Tong, H., \& Yuan, J. 2012, Astrophysics and Space Science,
  340, 307

\bibitem[{Wang {et~al.}(2000)Wang, Manchester, Pace, Bailes, Kaspi, Stappers,
  \& Lyne}]{wmp+00}
Wang, N., Manchester, R.~N., Pace, R., {et~al.} 2000, MNRAS, 317, 843

\bibitem[{{Watts} {et~al.}(2015){Watts}, {Espinoza}, {Xu}, {Andersson},
  {Antoniadis}, {Antonopoulou}, {Buchner}, {Datta}, {Demorest}, {Freire},
  {Hessels}, {Margueron}, {Oertel}, {Patruno}, {Possenti}, {Ransom}, {Stairs},
  \& {Stappers}}]{wxe+15}
{Watts}, A., {Espinoza}, C.~M., {Xu}, R., {et~al.} 2015, Advancing Astrophysics
  with the Square Kilometre Array (AASKA14), 43

\bibitem[{{Yu} {et~al.}(2013){Yu}, {Manchester}, {Hobbs}, {Johnston}, {Kaspi},
  {Keith}, {Lyne}, {Qiao}, {Ravi}, {Sarkissian}, {Shannon}, \& {Xu}}]{ymh+13}
{Yu}, M., {Manchester}, R.~N., {Hobbs}, G., {et~al.} 2013, MNRAS, 429, 688

\bibitem[{{Yuan} {et~al.}(2010){Yuan}, {Wang}, {Manchester}, \& {Liu}}]{ywml10}
{Yuan}, J.~P., {Wang}, N., {Manchester}, R.~N., \& {Liu}, Z.~Y. 2010, MNRAS,
  404, 289

\end{thebibliography}

\end{document}